\documentclass{article}
\usepackage{epsf,latexsym}
\usepackage[0]{metric}

\def\abstracts#1{\begin{abstract}#1\end{abstract}}
\let\STDauthor=\author
\def\author#1{\gdef\TMPauthor{#1}\STDauthor{#1}}
\def\address#1{\STDauthor{\TMPauthor\\#1}}

\newcounter {probnum}
\newenvironment{problem}%
   {\refstepcounter{probnum}\noindent \theprobnum.  }%
   {\vspace {0.25in}}

\newbox \JCCHoldBox
\newdimen \JCCLower
\newcommand \epscenterbox [2]%
{    
     \epsfxsize=#1%
     \setbox \JCCHoldBox = \hbox{\epsfbox{#2}}%
     \JCCLower = 0.5\ht\JCCHoldBox
     \advance \JCCLower by -0.3ex%
     \lower \JCCLower \box\JCCHoldBox
}

\def \MSbarbasic {\overline{{\rm MS}}}
\def \MSbar {\ifmmode \MSbarbasic \else $\MSbarbasic$\fi }
\def \dalem {\Box}
\def\half{{\textstyle \frac12}}
\def\quarter{{\textstyle \frac14}}

\def \st#1{\centeron{$#1$}{$/$}}
\def\centeron#1#2{{\setbox0=\hbox{#1}\setbox1=\hbox{#2}\ifdim
   \wd1>\wd0\kern.5\wd1\kern-.5\wd0\fi
   \copy0\kern-.5\wd0\kern-.5\wd1\copy1\ifdim\wd0>\wd1
   \kern.5\wd0\kern-.5\wd1\fi}}

\makeatletter
\def\THISFOOT#1{%
   \def \ps@THISFOOT {%
      \def \@oddfoot 
          {\parbox[t]{\textwidth}{\footnoterule \footnotesize #1}}%
      \let \@evenfoot \@oddfoot
   }
   \thispagestyle{THISFOOT}%
}
\makeatother

\begin{document}

\title {THE PROBLEM OF SCALES: RENORMALIZATION AND ALL THAT
}
\author {JOHN COLLINS}
\address{
    The Pennsylvania State University, 104 Davey Laboratory \\
    University Park PA 16802, U.S.A.}
\date{12 February 2009}

\maketitle

\THISFOOT{
   Published in  ``Theoretical Advanced Study Institute in
   Elementary Particle Physics, 1995: QCD and Beyond'', D.E. Soper,
   ed., (World Scientific Singapore);
   \texttt{http://arXiv.org/abs/hep-ph/951027}.  This version has
   corrections. 
   }

\abstracts{
    I explain the methods that are used in field theory for
    problems involving typical momenta on two or more widely
    disparate scales. The principal topics are: (a)
    renormalization, which treats the problem of taking an
    ultra-violet cut-off to infinity, (b) the renormalization
    group, which is used to relate phenomena on different scales,
    (c) the operator product expansion, which shows how to obtain
    the asymptotics of amplitudes when some of its external
    momenta approach infinity.
}


\section{Introduction}
\label{sec:intro}

There is a common principle in physics, and indeed in
science in general: One should neglect negligible
variables.  For example, a meteorologist does not have
to take in account the microscopic molecular structure
of air.  A chemist only needs to know the masses and
electric charges of atomic nuclei, but not their detailed
properties as given by QCD.  At the other end of the
scale, a particle physicist discussing collisions at
current energies need not take gravitation into account.

Without this principle, one would in general have great
difficulty in making the approximations that make theoretical
calculations tractable.  When discussing phenomena on
one scale of distance, one has no need to treat
phenomena on totally different scales, at least not in
any great detail.

Unfortunately, the principle, in its simplest form, does
not apply to quantum field theories.  Consider, for
example, the large $ \BDpos p^{2}$ limit of the propagator for a
scalar field in a renormalizable theory.  At lowest order
we have
\[\frac {i}{ \BDpos p^{2}-m^{2}}\simeq\frac {i}{ \BDpos p^{2}}\mbox{\rm \ when
$|p^{2}|/m^{2}\gg\infty $}
.\]
This suggests that the large $p^{2}$ asymptote of the full
propagator can be obtained simply by setting $m=0$, and
hence that by dimensional analysis the propagator is
proportional to $1/p^{2}$:
\[
   S(p^{2},m^{2})\simeq G(p^{2},0)
   = \frac {{\rm const}}{p^{2}} .
\]
As is well known, this is false, since higher order
corrections to the propagator are polynomial in $\ln p^{2}$ at
large $p^{2}$.  This anomalous scaling behavior is directly
associated with the need to renormalize ultra-violet
divergences in the theory.

So we find ourselves with a set of related topics that
form the subject of this set of lectures:
\begin{itemize}
\item\ Ultra-violet divergences and their
      renormalization.
\item\ Anomalous scaling in high momentum limits; the
      renormalization group.
\item\ Asymptotic behavior as some momenta get large;
      the operator product expansion (OPE).
\item\ How these ideas, after being developed in a simple
      theory $(\phi ^{4}$ theory), apply to the standard model
      and other field theories.
\end{itemize}
The concepts as expounded in these lectures apply to
situations that are formally characterized as ``Euclidean''.
Sterman's lectures will build on them to treat problems
of physical interest that are directly formulated in
Minkowski space.

I have been quite sparing of references, except for statements of
properties that I do not think are well-known.  I apologize to
authors whose work I have not recognized in the bibliography. The
reader shown consult the bibliography for more detailed
treatments, and for further references.


\section{Basic ideas of renormalization theory}
\label{sec:basics}

I will now summarize the basic ideas of renormalization
theory, first by expounding a well-known example from
classical physics and then by reviewing how
renormalization works at the one-loop order in simple
examples.

\subsection{Renormalization in classical physics}

Let $m_{0}$ be the bare mass of an electron, i.e., the mass of the
electron itself.  Then the total energy of the electron and of
its electric field down to radius $a$ is given by
\begin{eqnarray}
   mc^{2}&=&m_{0}c^{2}+\int _{|{\bf r}|>a}d^{3}{\bf r}\frac {e^{2}}{32\pi
^{2}\epsilon _{ 0}r^{4}}\nonumber\\
      &=&m_{0}c^{2}+\frac {e^{2}}{8\pi \epsilon _{0}a}.
\end{eqnarray}
The resulting linear divergence as $a\to 0$ is a consequence
of the assumption of a point electron and the
of special relativity, and was well-known in the early
part of this century.

In a consistent classical theory one would have to
introduce a structure for the electron and some
non-electromagnetic forces to hold the electric charge
together.  Then
\begin{equation}
   \mbox{\rm observed mass of electron}=
   \mbox{\rm bare mass}+
   \frac {\mbox{\rm energy in e-m and non-e-m fields}}{c^{2}}.
\end{equation}
The effect of the energy of the field on the
mass of the electron is very large: if we set $a=10^{-15}{\rm m}$,
then the electromagnetic energy is about $0.7\,{\rm MeV}$, rather
larger than the total energy.  We now know that the
electron is point-like to at least several orders of
magnitude smaller in distance, so the classical electromagnetic
energy is much larger than the total energy.  Of course, the
calculation above is purely classical, and it must be
modified by quantum field theoretical effects.  But it
illustrates some important principles.

One is that a basic parameter, such as a mass, does not
have to equal the observed quantity of the same name.

Another is that it is convenient to write all calculations in
terms of the measured mass $m$.  and then we write the bare mass
as the observed mass plus a counterterm:
\begin{equation}
   m_{0}=m+\delta m,
\end{equation}
with the counterterm being adjusted to cancel the
energy of the fields that surround an isolated electron.
This procedure we call renormalization of the mass.
At this point we may take the radius $a\to 0$, and we will
no longer encounter divergences in directly measurable
quantities.

{}From a practical point of view a very large finite value
for the self energy of an electron is almost as bad as an
infinity.  In either case, the field energy is a large
effect, and we must express calculations in terms of
measured quantities.

\subsection{Divergences in quantum field theory}

All the above issues, and in particular the divergences,
are even more pervasive in quantum field theory; there
are infinities everywhere, but they appear to be rather
more abstract in nature.

To explain the principles of how these divergences arise
and of how they are cancelled by renormalization
counterterms, I will use the simplest possible theory, $\phi ^{4}$
in four space-time dimensions.  Its Lagrangian density is
\begin{equation}
   {\cal L} = \BDpos \frac {1}{2}(\partial \phi )^{2}-\frac {1}{2}m^{2}\phi ^{ 2}-\frac
{1}{4!}\lambda \phi ^{4}.
\label{eq:L.original}
\end{equation}
The Feynman rules for the theory are given by a
propagator $i/( \BDpos p^{2}-m^{2}+i\epsilon )$ and a vertex $-i\lambda $.  Although we
will wish to treat the theory in four dimensions, it will
be convenient to ask what happens if the space-time
dimensionality had some other value $n$.  Dimensional
analysis will play an important role in our discussions,
so we note now that the field and coupling have energy
dimensions $[\phi ]=E^{n/2-1}$ and $[\lambda ]=E^{4-n}$.  These follow
from the fact that the units of ${\cal L}$ are those of an energy
density, since we use natural units, i.e., with
$\hbar =c=\epsilon _{0}=1$.

All the same principles will apply to more general theories, and
all theories in four-dimensional space-time have divergences, so
that we cannot evade the issues by trying another theory.
Indeed, it can readily be seen from other lectures in this book
that the topics I will discuss are essential in treatments of the
standard model and of its many proposed extensions.

\begin{figure}
    \begin{center}
    \leavevmode
       $\epscenterbox{0.6in}{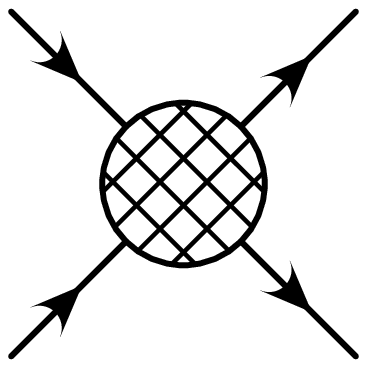}
        = \epscenterbox{0.4in}{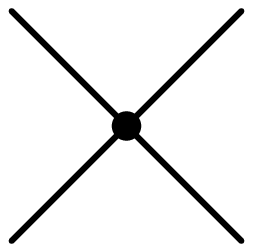}
        + \epscenterbox{0.6in}{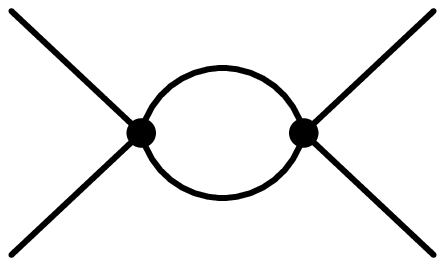}
        + \epscenterbox{0.6in}{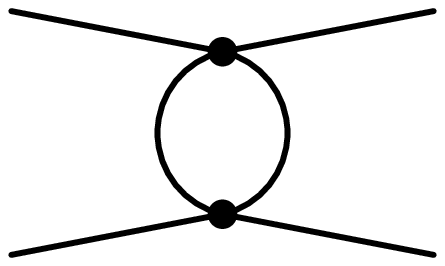}
        + \epscenterbox{0.6in}{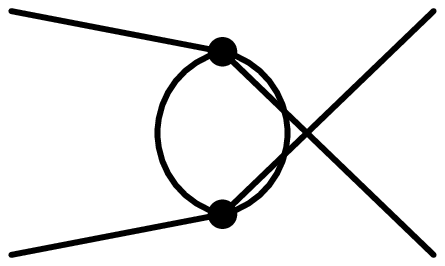}
        + O(\lambda ^{3}).
       $
    \end{center}
    \caption{Connected and amputated 4-point Green function
            in $\phi ^{4}$ theory.}
\label{fig:1loop4}
\end{figure}

The graphs up to one-loop order for the connected and
amputated 4-point Green functions are shown in
Fig.~\ref {fig:1loop4}.  Each of the one-loop graphs is the
same except for a permutation of the external lines, so
it will suffice to examine the integral
\begin{equation}
   I(p^{2})=\frac {(-i\lambda )^{2}}{2}\int \frac {d^{n}k}{(2 \pi )^{n}}\frac
{i^{2}}{( \BDpos k^{2}-m^{2}+i\epsilon )\,[\BDpos(p-k)^{2}-m^{2}+i\epsilon  ]}.
\label{eq:1loop}
\end{equation}
Possible divergences at the propagator poles are avoided
by contour deformations, according to the specified $i\epsilon $
prescriptions.  The only remaining divergence is an
``ultra-violet'' divergence as $k\to \infty $.  At the physical
space-time dimension, $n=4$, this is a logarithmic
divergence.

In general, we define the {\it degree of divergence}, $\Delta $, for a
graph by counting powers of $k$ at large $k$.  For Eq.\
(\ref{eq:1loop}) this is $\Delta = n-4$. Evidently, the graph is
convergent if $\Delta <0$, i.e., if $n<4$, and it is convergent if
$\Delta \geq 0$, i.e., if $n\geq 4$.

A convenient (and in fact general) way of seeing how
the divergence arises is to observe that the degree of
divergence is equal to the energy dimension of the
integral over $k$.  Since the one-loop graph has the same
dimension as the lowest order graph, whose value is $-i\lambda $,
we have
\begin{equation}
    \mbox{dim(tree graph)} = {\rm dim\,} \lambda
    = \mbox{dim(loop graph)}
    = {\rm dim\,} \lambda ^{2} + \Delta ,
\end{equation}
so that
\begin{equation}
   \Delta  = -{\rm dim\,} \lambda  = n-4.
\end{equation}
Thus the existence of an ultra-violet divergence is
directly associated with the fact of having a coupling of
zero or negative dimension.

\subsection{Interpretation of divergence}

Since we have obtained an infinite result for the
calculation, the theory is no good, at least as
formulated na\"\i vely.  (The fault is not just that of
perturbation theory, although that is not apparent at
this stage.)  Ultra-violet divergences are typical properties of
relativistic quantum field theories and are particularly
associated with the need for a continuous manifold for space-time
and for local interactions in a relativistic theory.

To remove the divergences, we first impose a ``cut-off'' on the
theory, to regulate the ultra-violet divergences.  This typically
removes some desirable properties: the use of a lattice
space-time, for example, wrecks Poincar\'e invariance.  Then we
adjust the parameters ($\lambda $, $m$) as functions of the cut-off, so
that finite limits are obtained for observable quantities when we
finally remove the cut-off.

This is an obvious generalization of the procedure for the
classical electron.  Our formulation is non-perturbative, even
though typical applications are within perturbation theory.  If
one does not like even the implicit divergences, then one need
not take the limit of completely removing the cut-off, but only
go far enough that the effect of the cut-off, after
renormalization, is negligible in actual experiments.

\subsection{Renormalization of one-loop graph}
\label{sec:ren1loop}

Suppose that we have a lattice for space or space-time, with
lattice spacing $a$.  Then the free propagator $S(k,m;a)$
approaches its continuum value
$i/( \BDpos k^{2}-m^{2}+i\epsilon )$ when $|k^{\mu }| \ll 1/a$, while it is zero if
$|k^{\mu }| > O(1/a)$.  The regulated graph Eq.\ (\ref{eq:1loop}) can be
written
\begin{eqnarray}
   I(p^{2})&=&\frac {(-i\lambda )^{2}}{32\pi ^{4}} \int  d^{4}k \, S(k,m;a)
S(p-k,m;a)
\nonumber\\
     &\simeq& \frac {(-i\lambda )^{2}}{32\pi ^{4}}
         \Bigg\{
           \int _{|k|<k_{0}} d^{4}k \, \frac {i^{2}}{( \BDpos k^{2}-m^{2}+i\epsilon
)\,[\BDpos(p-k)^{2}-m^{2}+i\epsilon  ]}.
\nonumber\\
&&~~~~~~~~~~~
           + \int _{|k|>k_{0}} d^{4}k \, S(k,0;a)^{2}.
        \Bigg\}
\label{eq:1loop.a}
\end{eqnarray}
Here we separate small and large momenta by a parameter
$k_{0}$ that obeys $p, m \ll k_{0} \ll 1/a$.  Since the divergent part of
the integral does not depend on the external momentum $p$, we can
cancel the divergence by a ``counterterm'' that corresponds to a
renormalization of the $\phi ^{4}$ interaction vertex:
\begin{eqnarray}
\lefteqn{
   \left[~
      \epscenterbox{0.3in}{4tree.eps}
      + \epscenterbox{0.3in}{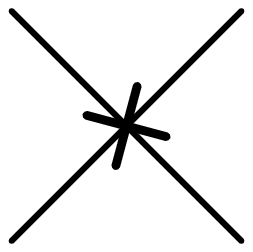}
   ~\right]
   + \epscenterbox{0.6in}{4loops.eps}
   + \epscenterbox{0.6in}{4loopt.eps}
   + \epscenterbox{0.6in}{4loopu.eps}
   + O(\lambda ^{3}).
}
\nonumber\\[0.1in]
   &=& -i\lambda  +
      i\lambda ^{2}\left\{\mbox{Complicated piece independent of $a$}
              + C(a) + O(|pa|)
        \right\}
\nonumber\\
&&
     -i\Delta \lambda
     +O(\lambda ^{3}),
\end{eqnarray}
where $C(a)$ diverges as $a\to 0$.  If we set $\delta \lambda =C(a)+{\rm
finite}$, we will get a finite result in the limit $a\to 0$ with the
renormalized coupling $\lambda $ fixed.

\subsection{Coordinate space}
\label{sec:CoordSpace}

(See \cite{JCC} for the material in this section.)
The same loop graph as before can be written in coordinate space
as
\begin{equation}
   \epscenterbox{1.0in}{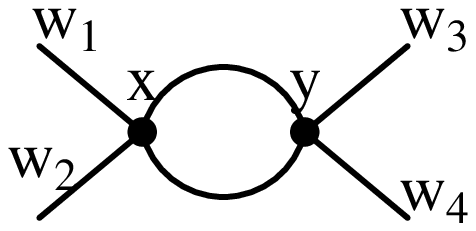}
   =
   \frac {(-i\lambda )^{2}}{2} \int  d^{4}x \, d^{4}y \, \left[ i\Delta
_{F}(x-y) \right]^{2}
   f(x,y, w_{1},w_{2},w_{3},w_{4}),
\end{equation}
where the propagator is $i\Delta _{F}$, and the function $f$ is the
product of the four external propagators.  Since
\begin{equation}
   i\Delta _{F}(x-y) \sim \frac {\BDneg 1}{4\pi ^{2} (x-y)^{2}}
\end{equation}
as $x\to y$, there is a logarithmic divergence at $x=y$, and
we can cancel the divergence by a counterterm proportional to
$\delta ^{(4)}(x-y)$.  Thus the value of the graph plus counterterm has
the form
\begin{eqnarray}
   \lim_{a\to 0} \frac {(-i\lambda )^{2}}{2} \int  d^{4}x \, d^{4}y \,
   \left\{\left[ i\Delta _{F}(x-y) \right]^{2} + C(a) \delta ^{(4)}(x-y)
   \right\}
   f(x,y, \underline{w}),
\end{eqnarray}
where I have indicated the limit that an ultra-violet regulator
is removed.

\subsection{Modified Lagrangian}

Our discussions indicate that we should attempt to cancel the UV
divergences by renormalizing all the parameters in the
Lagrangian.  Thus we replace the original Lagrangian Eq.\
(\ref{eq:L.original}) by
\begin{equation}
   {\cal L} = \BDpos \frac {1}{2} Z (\partial \phi )^{2} - \frac {1}{2}m_{B}^{2}\phi
^{2} - \frac {1}{4!}\lambda _{B}\phi ^{4},
\label{eq:LB}
\end{equation}
where $Z$, $m_{B}^{2}$ and $\lambda _{B}$ are singular as the lattice spacing
is
taken to zero.  By changing variables according to $\phi_{0}=\sqrt Z \phi $,
$m_{0}^{2}=m_{B}^{2}/Z$, and $\lambda _{0}=\lambda _{B}/Z^{2}$, we find
\begin{equation}
   {\cal L} = \BDpos \frac {1}{2}(\partial \phi_{0})^{2} - \frac {1}{2}m_{0}^{2}\phi
_{0}^{2} - \frac {1}{4!}\lambda _{0}\phi _{0}^{4},
\label{eq:L0}
\end{equation}
where the ``kinetic energy term'' has a unit coefficient.
We call $\phi _{0}$, $m_{0}$ and $\lambda _{0}$ the bare field, bare mass and
bare
coupling.

For doing perturbation theory, we separate ${\cal L}$ into three
pieces:
\begin{eqnarray}
   {\cal L} &=& \BDpos \frac {1}{2} (\partial \phi )^{2} - \frac {1}{2}m^{2}\phi ^{2}
\nonumber\\
   &&
    - \frac {1}{4!}\lambda \phi ^{4}
\nonumber\\
   &&
    \BDpos \frac {1}{2} \delta Z (\partial \phi )^{2} - \frac {1}{2}\delta m^{2}\phi
^{2} - \frac {1}{4!}\delta \lambda \phi ^{4} .
\label{eq:LR}
\end{eqnarray}
Here we have chosen arbitrarily a renormalized mass $m$ and a
renormalized coupling $\lambda $ that are to be held fixed as the UV
regulator is removed. The first line, the ``free Lagrangian'', is
used to make the propagators in the Feynman rules, and the
remaining terms are used to make the interactions, to be treated
in perturbation theory.  The second line gives the ``basic
interaction'' and the third line, the ``counterterm Lagrangian''
is used to cancel divergences in graphs containing basic
interaction vertices.  The counterterms are expressed in terms of
the renormalized parameters, and the renormalized coupling $\lambda $ is
used as the expansion parameter of perturbation theory.

It is to be emphasized that the three formulae
(\ref{eq:LB}--\ref{eq:LR}), refer to the same Lagrangian, which
is just written differently.  Some people talk about a different
bare and renormalized Lagrangian, but this is not the case in the
way I have formulated renormalization.  The one Lagrangian that
is different, Eq.\ (\ref{eq:L.original}) is not actually used for
physical calculations; indeed it gives unambiguously divergent
Green functions.  I will sometimes refer to the Eq.\
(\ref{eq:L0}) as the ``bare Lagrangian'', but what I will
actually mean is the ``Lagrangian written in terms of bare
fields''.

\subsection{Dimensional regularization}
\label{sec:Dim.Reg}

With suitable adjustments of the finite parts of the
counterterms, the same results are obtained for renormalized
Green functions (i.e., Green functions of the renormalized field
$\phi $) no matter which UV regulator is chosen.  Hence we can use
the regulator that is most convenient for doing calculations.

For many purposes, dimensional regularization is very convenient.
Here, we treat the dimension $n$ of space-time as a continuous
parameter, $n=4-\epsilon $.  We remove the regulator by taking $\epsilon \to
0$.  It
can be shown (e.g., \cite{JCC}) that the integrals over $n$
dimensional vectors can be consistently defined, so that
dimensional regularization is consistent, at the level of
perturbation theory.

The one complication is that the coupling has mass dimension $\epsilon $.
To make calculations manifestly dimensionally
correct, we introduce an arbitrary parameter with the dimensions
of mass, the ``unit of mass'' $\mu $ and write the bare coupling as
\begin{equation}
   \lambda _{0} = \mu ^{\epsilon }\lambda  + \mbox{counterterms},
\end{equation}
where now the renormalized coupling is dimensionless for all
$\epsilon $.\footnote{The reader will probably notice that there is a
   considerable arbitrariness in the conversion of the Lagrangian
   in its bare form Eq.\ (\ref{eq:L0}) to Feynman rules.  A change
   of the value of $\mu $ changes the split of the Lagrangian
   between free, basic and counterterm parts.  As we will see in
   our discussion of the renormalization group, Sect.\ \ref{sec:RG},
   such changes can be compensated by changes in the values of
   the renormalized parameters.  Instead of being a nuisance,
   they give useful methods of optimizing the information
   obtained by finite order calculations.
}
The one-loop integral Eq.\ (\ref{eq:1loop}) is now
\begin{equation}
   I(p^{2})=\frac {\lambda ^{2} \mu ^{2\epsilon }}{2(2\pi )^{4-\epsilon }}
\int  d^{4-\epsilon }k  \frac {1}{( \BDpos k^{2}-m^{2}+i\epsilon
)\,[\BDpos(p-k)^{2}-m^{2}+i\epsilon  ]}.
\label{eq:1loop.dim.reg}
\end{equation}

Calculation and renormalization of Feynman graphs become easy
with dimensional regularization.  For example, in Eq.\
(\ref{eq:1loop.dim.reg}), we combine the denominators by the
Feynman parameter method Eq.\ (\ref{eq:Feyn.param}), shift the
integral over $k$ and then perform a Wick rotation: $k^{0}=i\omega $ to
obtain a spherically symmetric integral over a Euclidean
momentum:
\begin{equation}
   I(p^{2}) = \frac {i \lambda ^{2} \mu ^{2\epsilon }}{2(2\pi )^{4-\epsilon }}
\int _{0}^{1}dx \int  d_{E}^{4-\epsilon }k
           \frac {1}{\left[ m^{2} \BDminus p^{2}x(1-x) + k_{E}^{2} \right]^{2}},
\end{equation}
where $d_{E}^{4-\epsilon }k = d\omega \, d^{3-\epsilon }{\bf k}$, and
$k_{E}^{2} = \omega ^{2}+{\bf k}^{2} = \BDneg k^{2}$.  It can readily be shown that a
spherically symmetric integral can be reduced to a radial
integral Eq.\ (\ref{eq:sph.sym}), and hence we get
\begin{eqnarray}
   I(p^{2}) &=& \frac {i \lambda ^{2} \mu ^{\epsilon }}{32\pi ^{2}} \Gamma
(\epsilon /2)
            \int _{0}^{1}dx \left[  \frac {m^{2} \BDminus p^{2}x(1-x)}{4\pi \mu ^{2}}
\right]^{-\epsilon /2}
\nonumber\\
   &=& \frac {i \lambda ^{2} \mu ^{\epsilon }}{32\pi ^{2}}
      \left\{
         \frac {2}{\epsilon }
         -\int _{0}^{1}dx \,
           \ln\left[  \frac {m^{2} \BDminus p^{2}x(1-x)}{4\pi \mu ^{2}} \right]
         -\gamma _{E}
         +O(\epsilon )
      \right\},
\end{eqnarray}
where $\gamma _{E}=0.5772\dots$ is Euler's constant.
In obtaining the first line of this equation, we have in effect
proved Eq.\ (\ref{eq:MSI0}).

We can now choose the counterterm
\begin{equation}
   \delta \lambda  = \frac {3\lambda ^{2}\mu ^{\epsilon }}{16\pi ^{2}\epsilon }
+ O(\lambda ^{3})
\label{eq:CT3b}
\end{equation}
to cancel the divergence in the three one-loop graphs.  Then
after adding the graphs and the counterterms we set $\epsilon =0$ and
obtain the renormalized amputated four-point Green function:
\begin{equation}
   -i\lambda
   - \frac {i \lambda ^{2}}{32\pi ^{2}} \sum _{p^{2}=s,t,u}
        \left\{
           \int _{0}^{1}dx \ln\left[  \frac {m^{2} \BDminus p^{2}x(1-x)}{4\pi \mu ^{2}}
\right]
           + \gamma _{E}
        \right\}
   +O(\lambda ^{3}).
\label{eq:1loop.R}
\end{equation}
Given our choice of the counterterm which is just a pure pole at
$\epsilon =0$, we call $\lambda $ the renormalized coupling in the minimal
subtraction scheme.

\subsection{Renormalizability}

We have shown that particular UV divergences in quantum field
theory can be removed by renormalization of the parameters of the
lagrangian.  The question now arises as to which theories can
have all their divergences renormalized away, a question which we
will address in Sects.\ \ref{sec:all.orders} and
\ref{sec:other.theories}.  It is convenient to make the following
definitions:
\begin{itemize}
   \item[1.]  A quantum field theory is {\em renormalizable} if
   it can be made finite as the UV regulator is removed by
   suitably adjusting coefficients in ${\cal L}$.

   \item[2.]  A quantum field theory is {\em non-renormalizable}
   if it is not renormalizable.  (To be strictly correct, I
   should say that in practice one says that a theory is
   non-renormalizable if it merely has not been shown to be
   renormalizable.  For example, proofs within perturbation
   theory do not automatically imply anything about
   non-perturbative properties of a theory.)

\end{itemize}
Typically proofs are in perturbation theory, to all orders.

\subsection{Divergent (one-loop) graphs in $(\phi ^{4})_{4}$}

An important element of proofs of renormalizability is the
concept of the ``degree of divergence'', $\Delta (\Gamma )$, of a graph $\Gamma
$.
As I have already said, this is defined by counting the number of
powers of loop momenta in the graph when these loop momenta get
large. A one-loop graph $\Gamma $ is UV divergent if $\Delta (\Gamma )\geq 0$
and UV
convergent if $\Delta (\Gamma )<0$. One should only include the
one-particle-irreducible (1PI) part of the graph in this
counting.  The same concept will be applied to higher-order
graphs in Sect.\ \ref{sec:all.orders}.

In $\phi ^{4}$ theory in 4 dimensions, one can easily show that, for a
general 1PI graph
\begin{equation}
   \Delta  = 4 - \mbox{number of external lines}.
\label{eq:DegDivPhi4}
\end{equation}
Hence the 4-point function is logarithmically divergent, and the
only other graphs that have a divergence are the self-energy
graphs, with $\Delta =2$.  By differentiating self-energy graphs three
times with respect to the external momenta we obtain convergent
integrals, with $\Delta =-1$.  The necessary counterterms are therefore
quadratic in external momenta; given Lorentz invariance, they
have the form $\BDpos i\delta Zp^{2}-i\delta m^{2}$. Corresponding vertices are
obtained
from ``wave-function'' and mass counterterms in the Lagrangian:
$\BDpos \frac {1}{2} \delta Z (\partial \phi )^{2} - \frac {1}{2}\delta m^{2}\phi
^{2}$, as in Eq.\ (\ref{eq:LR}). It now follows
that $\phi ^{4}$ theory is renormalizable at the one-loop level.  (In
fact the topology of the one-loop self-energy graph makes it have
no momentum dependence, so that $\delta Z$ starts at the next order,
$\lambda ^{2}$.)

\subsection{Summary}

\noindent 1.  A parameter in the Lagrangian or Hamiltonian need
not equal a measured quantity of the same name.

\noindent 2.  We can use simple power counting methods to
determine a degree of divergence for Feynman graphs, and hence
readily determine which are divergent.

\noindent 3.  We can cancel divergences by renormalization --
adjustment of the values of parameters in the Lagrangian compared
to the values one first thought of.

\noindent 4.  Renormalization can be implemented in perturbation
calculations by a counterterm technique.

\noindent 5.  We have reviewed some practical methods for
calculations, including dimensional regularization.



\section{All-orders Renormalization}
\label{sec:all.orders}

In this section I will summarize how one generalizes the one-loop
results of the previous section to all orders of perturbation
theory.  This subject has a reputation for being very abstruse.
To make it as comprehensible as possible, proper organization is
vital.  The distribution-theoretic methods developed by Tkachov
and collaborators \cite{Tkachov} are very useful, and my
presentation will use them.

I will also work in coordinate space.  This makes the
demonstrations more intuitive than the corresponding ones in
momentum-space.  Of course a theorem about Feynman graphs in
coordinate space immediately implies the corresponding theorem in
momentum space, which is where practical calculations are
normally done.

My starting point is the decomposition Eq.\ (\ref{eq:LR}) of the
Lagrangian into a free Lagrangian, a basic interaction term and
counterterms.  Each counterterm will be treated as a sum of
terms, one for each ``overall divergent'' graph generated by the
basic interaction.

We organize perturbation theory in powers of the renormalized
coupling by first writing the graphs of some given order that are
obtained by using only the basic interaction to form vertices;
these graphs have divergences.  Then to each such basic
graph we add all possible counterterm graphs; these are obtained
by replacing particular subgraphs by the corresponding
counterterm vertices in the counterterm Lagrangian.  The total of
each basic graph and its associated counterterm graphs will be
finite if the renormalization procedure works.

\begin{figure}
    \begin{center}
    \leavevmode
       $ \epscenterbox{0.6in}{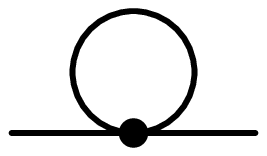}
       + \epscenterbox{0.6in}{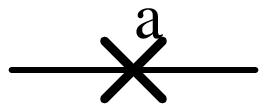}
       $
    \end{center}
    \caption{Renormalized self energy graph.}
\label{fig:1loop2R}
\end{figure}

\begin{figure}
    \begin{center}
    \leavevmode
        $\epscenterbox{0.8in}{4loops.eps}
        + \epscenterbox{0.8in}{4loopt.eps}
        + \epscenterbox{0.8in}{4loopu.eps}
        + \epscenterbox{0.5in}{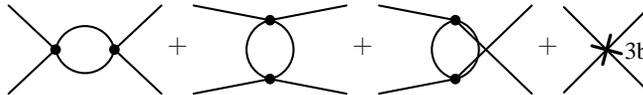}
        $
    \end{center}
    \caption{Renormalized 1-loop four-point function.}
\label{fig:1loop4R}
\end{figure}

Examples of this organization are shown in Figs.\
\ref{fig:1loop2R} and \ref{fig:1loop4R} for the $\phi ^{4}$ theory.  In
these figures we label the counterterms for each divergent basic
graph by a letter.  For example, in Fig.\ \ref{fig:1loop2R},
``$a$'' represents the one-loop mass counterterm
\begin{equation}
   -i\delta m^{2}_{a} = -i\frac {\lambda  m^{2}}{16\pi ^{2}\epsilon },
\label{eq:CTa}
\end{equation}
which can be calculated by the same methods we used in Sect.\
\ref{sec:Dim.Reg}.  In Fig.\ \ref{fig:1loop4R}, the counterterm
``$3b$'' is the one in Eq.\ (\ref{eq:CT3b}); the ``$3$'' in
``$3b$'' is an indication that there are three basic graphs, each
with a numerically identical counterterm.

\subsection{Dimensionally regularization and coordinate space}

In Sect.\ \ref{sec:CoordSpace}, I summarized how to renormalize
the one-loop four-point function in coordinate space.  Now I will
give some details that are needed to be able to do these
calculations in detail, with dimensional regularization.  The
free propagator and its asymptotic short-distance behavior are:
\begin{eqnarray}
   i\Delta _{F}(x-y) &=& \frac {m^{n/2-1}}{(2\pi )^{n/2}}
                \left[ \BDneg(x-y)^{2} \right]^{1/2-n/4}
                K_{n/2-1} \left( \sqrt {\BDneg m^{2}(x-y)^{2}} \right)
\nonumber\\
        &\sim& \frac {\Gamma (n/2-1)}{4\pi ^{n/2} \left( \BDneg (x-y)^{2}
\right)^{n/2-1}}
        ~~~\hbox{as $x\to y$}.
\label{eq:DeltaF}
\end{eqnarray}
Here $K_{n/2-1}$ is a modified Bessel function of order $n/2-1$.
Then the dimensionally regularized value of one of the one-loop
graphs, plus its counterterm, is
\begin{equation}
   \frac {(-i\lambda )^{2} \mu ^{\epsilon }}{2} \int  d^{n}x \, d^{n}y \,
   f(x,y, \underline{w})
   \left\{
      \mu ^{\epsilon } \left[ i\Delta _{F}(x-y) \right]^{2}
      +\frac {i}{8\pi ^{2}\epsilon } \delta ^{(n)}(x-y)
   \right\}.
\label{eq:Ren4}
\end{equation}

It should be noted that the limit as $\epsilon \to 0$ of Eq.\
(\ref{eq:Ren4}) can be considered as the integral of a test
function $f(x,y, \underline{w})$ with a generalized function.
The generalized function is the factor in braces ($\{\dots\}$),
and can be considered a generalization of the generalized
function $(1/x)_{+}$ in one dimension.

\subsection{List of results}

The mathematical problem to be solved is to understand the
asymptotics of the integration of multidimensional integrals in
momentum space as some or all of the integration variables go to
infinity.  Equivalently, it is to understand the short-distance
asymptotics of the corresponding coordinate-space integrals.

Our results will be
\begin{enumerate}
\item
    To associate a counterterm with each 1PI graph that has an
    overall divergence.

\item
    To show that the result is finite when we add to a basic
    graph the counterterms for its divergent subgraphs.

\item
    To obtain the power-counting techniques that determine the
    necessary counterterms.

\item
    To implement the counterterms as extra terms in the
    Lagrangian.

\end{enumerate}

\subsection{The Bogoliubov $R$-operation}

We define the renormalized value of a basic graph to be the sum
of the graph and all the counterterms for its divergences.  The
operation of constructing the renormalized value $R(\Gamma )$ of a
basic graph $\Gamma $ was formalized by Bogoliubov, hence the name
``Bogoliubov $R$-operation''.  An $N$-point Green function is
then to be considered as $\sum _{\Gamma } R(\Gamma )$, the sum being over all
basic
graphs for the Green function.

\begin{figure}
    \begin{eqnarray*}
    \lefteqn{
        \epscenterbox{1.0in}{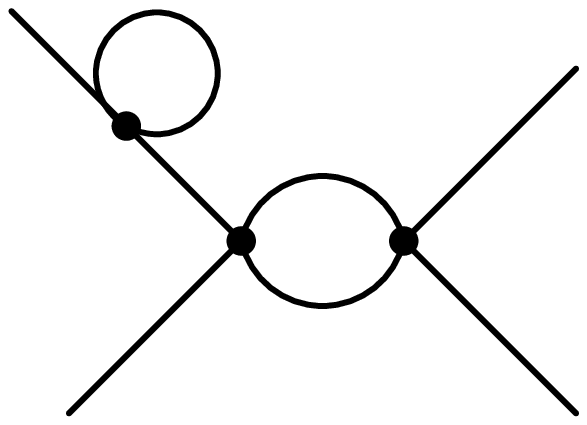}
        + \epscenterbox{1.0in}{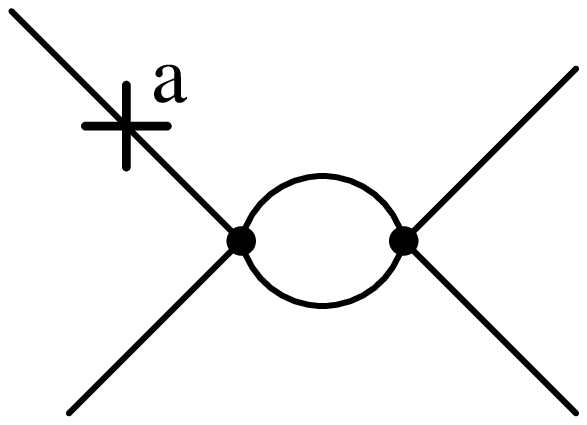}
        + \epscenterbox{0.8in}{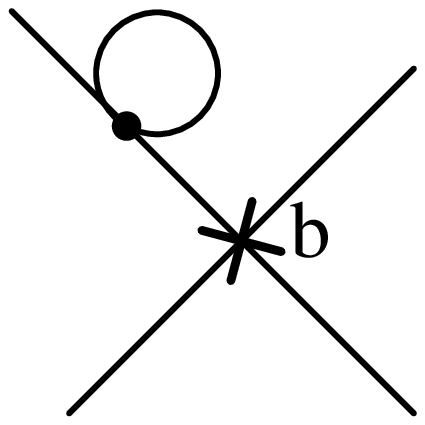}
        + \epscenterbox{0.8in}{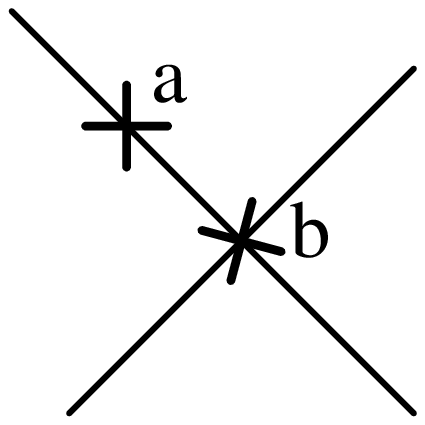}
    }
    \\[0.1in]
        &&
        \begin{array}{cccccc}
           &\epscenterbox{0.2in}{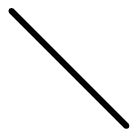}
        \\
           &&\left[ \epscenterbox{0.5in}{2loop.eps}
              + \epscenterbox{0.5in}{2tree-ct.eps}
             \right]
        \\
           &&&\epscenterbox{0.2in}{line-d.eps}
           &&\epscenterbox{0.2in}{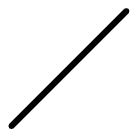}
        \\
           =&&&&
           \left[
              \epscenterbox{0.6in}{4loops.eps}
              + \epscenterbox{0.35in}{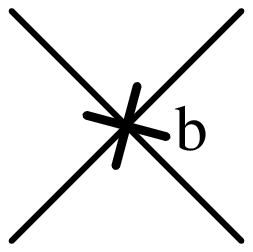}
           \right]
        \\
           &&&\epscenterbox{0.2in}{line-u.eps}
           &&\epscenterbox{0.2in}{line-d.eps}
        \end{array}
    \end{eqnarray*}
    \caption{An example of a renormalized two-loop graph.}
\label{fig:2loop1}
\end{figure}

Consider first a graph consisting of some one-loop 1PI graphs
connected together, as in Fig.\ \ref{fig:2loop1}.  The one-loop
counterterm vertices from  ${\cal L}$ imply the three counterterm
graphs in which one or both the divergent one-loop subgraphs is
replaced by its counterterm.  As shown in the lower part of the
figure, the counterterms may be organized to give a product form.

This example is evidently a case of a general set of results:
\begin{itemize}

\item
    UV divergences are confined to 1PI (sub)graphs.

\item
    Each divergent 1PI subgraph $\gamma $ ``owns'' a piece of
    counterterm $C(\gamma )$.

\item
    To see finiteness easily, we should keep a graph and its
    counterterm together.

\end{itemize}
Moreover we can easily generalize Fig.\ \ref{fig:2loop1} to the
following theorem:
\begin{quote}
    Suppose a basic graph $\Gamma $ is a product of lines outside loops
    and a set of 1PI graphs:
    \begin{equation}
      \Gamma  = \prod \mbox{propagators outside loops} ~~\times ~~
          \prod \mbox{1PI $\gamma $} .
    \end{equation}
    Then its renormalized value is
    \begin{equation}
      R(\Gamma ) = \prod \mbox{propagators outside loops} ~~\times ~~
          \prod R(\gamma ) .
    \end{equation}
\end{quote}

\subsubsection{Simple non-trivial case}

\begin{figure}
    \begin{center}
        \leavevmode
        \epscenterbox{1.2in}{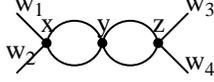}
    \end{center}
    \caption{A two-loop graph.}
\label{fig:2loop2}
\end{figure}

The basic graph in Fig.\ \ref{fig:2loop2} is
\begin{equation}
   \frac {i\lambda ^{3} \mu ^{3\epsilon }}{4} \int  d^{n}x \, d^{n}y \, d^{n}z
\,
   f(x,y,z, \underline{w})
   \left[ i\Delta _{F}(x-y) \right]^{2}
   \left[ i\Delta _{F}(y-z) \right]^{2} ,
\label{eq:2loop2}
\end{equation}
where $f(x,y,z, \underline{w})$ is the product of the four
external propagators $i\Delta _{F}(w_{1}-x) i\Delta _{F}(w_{2}-x) i\Delta
_{F}(w_{3}-z) i\Delta _{F}(w_{4}-z)$.
The singularities of the integrand are on the hyperplanes $x=y$
and $z=y$ and at their intersection $x=y=z$.
We ignore for this purpose the {\em integrable} singularities of
the four external propagators.

\begin{figure}
    \begin{center}
    \leavevmode
        $\epscenterbox{1.0in}{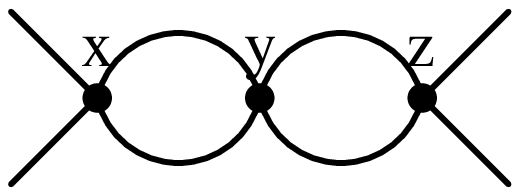}
        ~+~ \epscenterbox{0.8in}{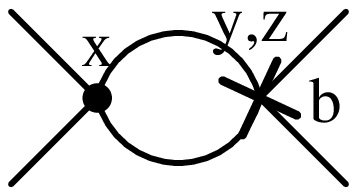}
        ~+~ \epscenterbox{0.8in}{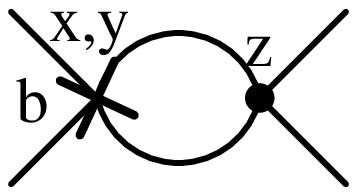}
        $
    \end{center}
    \caption{The two-loop graph of Fig.\ \protect\ref{fig:2loop2}
         plus counterterms for its subdivergences.}
\label{fig:2loop2Rprime}
\end{figure}

The Feynman rules tell us that we have counterterms for the
subdivergences, i.e., for the singularities on each of the
hyperplanes $x=y$ and $y=z$.  After adding these counterterms, as
in Fig.\ \ref{fig:2loop2Rprime}, we get
\begin{eqnarray}
\lefteqn{
   \frac {i \lambda ^{3} \mu ^{\epsilon }}{4} \int  d^{n}x \, d^{n}y \, d^{n}z
\,
   f(x,y,z, \underline{w})
   \times
}
\nonumber\\
&& \hspace*{-0.05in}
   \Big\{
      \mu ^{\epsilon } \left[ i\Delta _{F}(x-y) \right]^{2}
      \mu ^{\epsilon } \left[ i\Delta _{F}(y-z) \right]^{2}
\label{eq:2loop2Rprime}
\\
&&
      + \mu ^{\epsilon } \left[ i\Delta _{F}(x-y) \right]^{2}
        i C(\epsilon ) \,  \delta ^{(n)}(y-z)
      + i C(\epsilon ) \,  \delta ^{(n)}(x-y)
        \mu ^{\epsilon } \left[ i\Delta _{F}(y-z) \right]^{2}
   \Big\} ,
\nonumber
\end{eqnarray}
where $C(\epsilon ) = 1/(8\pi ^{2}\epsilon )$ is the same counterterm
coefficient as in
Eq.\ (\ref{eq:Ren4}).  When we take $\epsilon \to 0$, the integral is finite
everywhere except at the intersection of the singular surfaces of
the subdivergences, i.e., at $x=y=z$.  Thus we have subtracted
the subdivergences.
Therefore we make the following definitions:
\begin{itemize}

\item
    A {\em subdivergence} is where the positions of a strict
    subset of vertices approach each other and give a divergence.

\item
    An {\em overall divergence} occurs where the positions of all
    the vertices approach each other and give a divergence.

\end{itemize}
For example, Fig.\ \ref{fig:2loop2} has 2 subdivergences, at
$x=y$ and $y=z$ and has an overall divergence at $x=y=z$.

\begin{figure}
    \begin{center}
        \leavevmode
        \epscenterbox{1.1in}{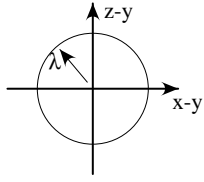}
    \end{center}
    \caption{Definition of radial and angular variables. }
\label{fig:lambda}
\end{figure}

A general method of finding the overall divergence is illustrated
by this graph.  Fix $y$, and consider the integral over $x-y$ and
$z-y$; this will be sufficient to find the divergences.  Now
express the integration variables in terms of a radial variable
$\lambda $ and some angular integrals over a $7-2\epsilon $ surface surrounding
the point $z-y=0=x-y$, Fig.\ \ref{fig:lambda}.  We write
\begin{equation}
   (x-y, \, z-y) = \lambda  (\hat x - \hat y, \, \hat z - \hat y ),
\end{equation}
so that
\begin{equation}
   \int  d^{n}x \, \int  d^{n}z = \int  d\lambda  \, \lambda ^{2n-1}
    \int  d^{2n-1}(\hat x - \hat y, \, \hat z - \hat y ) .
\end{equation}
The integral over the hatted variables gives no divergence,
because of the subtractions in Eq.\ (\ref{eq:2loop2Rprime}), and
simple power counting then shows that there is a logarithmic
divergence at $\lambda =0$.\footnote{
   The subtracted logarithmic singularities of the integrand at
   $x=y$ and $z=y$ include a term proportional to $\ln \lambda $ when the
   positions are scaled by a factor $\lambda $.  This logarithm does not
   affect the counting of powers, of course.
}

\begin{figure}
    \begin{center}
    \leavevmode
        $\epscenterbox{1.1in}{4loop2.eps}
        ~+~ \epscenterbox{0.7in}{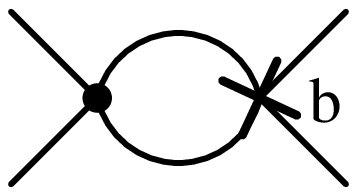}
        ~+~ \epscenterbox{0.7in}{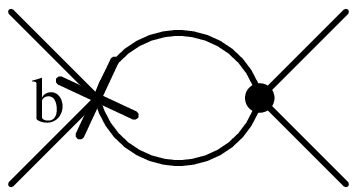}
        ~+~ \epscenterbox{0.3in}{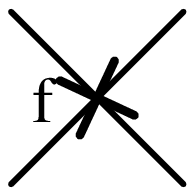}
        $
    \end{center}
    \caption{Fig.\ \protect\ref{fig:2loop2} plus all its
       counterterms.}
\label{fig:2loop2ct}
\end{figure}

Since the divergence is logarithmic, it may be subtracted by a
delta-function counterterm, as in Fig.\ \ref{fig:2loop2ct}, so
that the renormalized value of the graph is
\begin{eqnarray}
\lefteqn{
   \frac {i \lambda ^{3} \mu ^{\epsilon }}{4} \int  d^{n}x \, d^{n}y \, d^{n}z
\,
   f(x,y,z, \underline{w})
   \times
}
\nonumber\\
&&~~~
   \Big\{
      \mbox{Same as in Eq.\ (\protect\ref{eq:2loop2Rprime})}
      + C_{f}(\epsilon ) \,  \delta ^{(n)}(x-y) \, \delta ^{(n)}(z-y)
   \Big\} ,
\label{eq:2loop2R}
\end{eqnarray}
where calculation gives
\begin{equation}
   C_{f}(\epsilon ) = \frac {-1}{(8\pi ^{2}\epsilon )^{2}} .
\end{equation}

\subsection{The $R$-operation}

The structure of the previous example is quite general.\footnote{
  It in fact gives the simplest example of what is called an
  ``overlapping divergence''.  Such divergences were considered a
  difficult problem in the early days of renormalization theory,
  but with proper organization they are no more difficult to
  handle than non-overlapping divergences in multi-loop graphs.
}
The general procedure for obtaining the renormalized value $R(G)$
of a Feynman graph $G$ is summarized in the following formula:
\begin{equation}
   R(G) = G + \sum _{\gamma _{1},\dots,\gamma _{n}} G \Bigg|_{\gamma _{i}\to
C(\gamma _{i})} .
\label{eq:RDef}
\end{equation}
The sum is over all sets of non-intersecting 1PI subgraphs of
$G$, and each of the 1PI subgraphs $\gamma _{i}$ is to be replaced by its
counterterm $C(\gamma _{i})$.  The counterterm $C(\gamma )$ of a 1PI graph
$\gamma $
is constructed in the form
\begin{equation}
   C(\gamma ) = - T \left(
                  \gamma  + \mbox{Counterterms for subdivergences}
              \right) .
\label{eq:CDef}
\end{equation}
Here $T$ is an operation that defines the renormalization scheme.
For example, in minimal subtraction we define
\begin{equation}
   T(\Gamma ) = \mbox{pole part at $\epsilon =0$ of $\Gamma $}.
\end{equation}
In Eq.\ (\ref{eq:CDef}), we can formalize the term inside square
brackets as follows:
\begin{eqnarray}
   \bar R (\gamma ) &\equiv & \gamma  + \mbox{Counterterms for subdivergences}
\nonumber\\
    &=& \gamma  + \mathop{\sum \nolimits'}\limits_{\gamma _{1},\dots,\gamma
_{n}} \gamma  \Bigg|_{\gamma _{i}\to C(\gamma _{i})} ,
\label{eq:RPrimeDef}
\end{eqnarray}
where the prime (${}'$) on the $\sum \nolimits'$ denotes that we sum
over all sets of non-intersecting 1PI subgraphs except for the
case that there is a single $\gamma _{i}$ equal to the whole graph (i.e.,
$\gamma _{1}=\gamma $). Thus
\begin{equation}
   R(\gamma ) = \bar R(\gamma ) + C(\gamma ) .
\end{equation}

The formulae Eq.\ (\ref{eq:RDef})--(\ref{eq:RPrimeDef}) form a
recursive construction of the renormalization of an arbitrary
graph.  The recursion starts on one-loop graphs, since
they have no subdivergences, i.e., $C(\gamma )=-T(\gamma )$ for a one-loop
1PI graph.

The important non-trivial analytic result, an example of
which we saw in the previous subsection, is that power counting
in its na\"{\i}vest form determines what counterterms are needed
for a given 1PI (sub)graph.  The counterterm $C(\gamma )$ is a
polynomial of degree equal to the overall degree of divergence of
$\gamma $.  Thus the na\"{\i}ve degree of divergence equals the actual
degree of divergence, but only after subtraction of
subdivergences.

Once the counterterms have been constructed, we may implement
them as counterterms in the Lagrangian.  Thus for the $\phi ^{4}$ theory
\begin{equation}
   -i \delta \lambda  = \sum  C(\mbox{1PI graph for 4-point function}) ,
\end{equation}
and
\begin{equation}
   \BDpos i \delta Z p^{2} - i \delta m^{2} = \sum  C(\mbox{1PI self-energy graphs}) .
\end{equation}

\begin{figure}
    \begin{center}
    \leavevmode
        $\epscenterbox{0.6in}{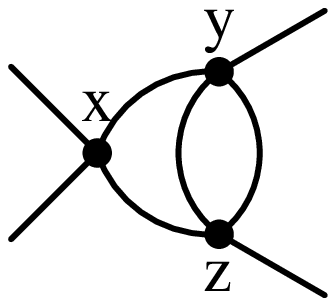}
        ~+~ \epscenterbox{0.8in}{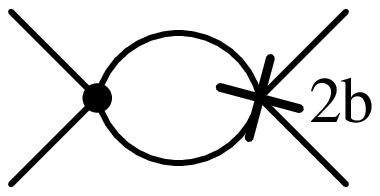}
        ~+~ \epscenterbox{0.3in}{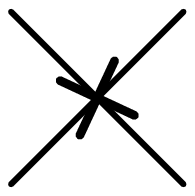}
        $
    \end{center}
    \caption{Another two-loop graph.}
\label{fig:2loop3}
\end{figure}

Another example of 2-loop renormalization is shown in Fig.\
\ref{fig:2loop3}.  The symmetry factors work out such that two
copies of the one-loop counterterm are needed for the
subdivergence.

\subsection{Translation to momentum space}

We may translate the above results to momentum space.  For
example, the 1PI part of the two-loop graph of Fig.\
\ref{fig:2loop2} in momentum space is
\begin{equation}
   \frac {i\lambda ^{3} \mu ^{3\epsilon }}{4 (2\pi )^{8-2\epsilon }} \int
d^{n}k \, d^{n}l \,
   \frac {1}{( \BDpos k^{2}-m^{2}) \, \left[ \BDpos (p-k)^{2} - m^{2} \right]}
   \frac {1}{(\BDpos l^{2}-m^{2}) \, \left[ \BDpos (p-l)^{2} - m^{2} \right]} .
\end{equation}
It has subdivergences when $k\to \infty $ with $l$ fixed, and when $l\to \infty
$
with $k$ fixed, and it has an overall divergence when both
$k,l\to \infty $.  The interplay between the subdivergences and the
overall divergence is harder to conceptualize when they are
at infinity, as they are in momentum space, than when they are
localized on lower dimensional surfaces and at points, as they are
in coordinate space.

But the calculations are quite easy, particularly for Fig.\
\ref{fig:2loop2}.  The basic graph is just the square of a simple
integral:
\begin{eqnarray}
\lefteqn{
   -i\lambda ^{3}\mu ^{\epsilon } \, I(p^{2},m^{2},\epsilon )^{2}
}
\label{eq:2loop3}
\\
   &=& -i\lambda ^{3}\mu ^{\epsilon }
     \left\{
       \frac {A}{\epsilon } - \frac {A}{2} \int _{0}^{1} dx \,
           \ln\left[
                \frac {m^{2} \BDminus p^{2}x(1-x)}{\mu ^{2}}
              \right]
       + \mbox{constant}
       + O(\epsilon )
     \right\}^{2} ,
\nonumber
\end{eqnarray}
where $A=1/(16\pi ^{2})$.
This integral has a double pole ($\propto 1/\epsilon ^{2}$) at $\epsilon =0$,
which is
evidently a characteristic of two loop graphs with a divergent
subgraph inside the overall divergence.  But the single pole
($\propto 1/\epsilon $) has a non-polynomial coefficient.  Hence, before we
subtract subdivergences we cannot renormalize the graph.

After subtraction of subdivergences, the only divergences are
independent of the external momentum, as befits a logarithmic
divergence:
\begin{eqnarray}
   \bar R(\mbox{Fig.\ \protect\ref{fig:2loop3}})
   &=&
   -i\lambda ^{3}\mu ^{\epsilon } \left( I^{2} - 2 I \times  \mbox{pole}(I)
\right)
\nonumber\\
   &=&
   -i\lambda ^{3}\mu ^{\epsilon } \left[ \left( I - \mbox{pole}(I) \right)^{2}
                 - \mbox{pole}(I)^{2}
          \right]
\nonumber\\
   &=&
   -i\lambda ^{3}\mu ^{\epsilon } \left( \mbox{finite} - \frac {A^{2}}{\epsilon
^{2}} \right) .
\label{eq:2loop3.Rbar}
\end{eqnarray}
Hence the graph can be renormalized by a coupling counterterm, the
same one as we calculated by coordinate space methods.

Note that at large $p^{2}$, the finite part behaves like $\ln^{2}p^{2}$
plus smaller terms, whereas a one-loop graph only has one
logarithm.  The double logarithm is characteristic of two-loop
graphs with subdivergences.

\subsection{Summary}

\noindent 1.  Suppose we have added to the Lagrangian a
counterterm $C(\gamma )$ for a particular graph $\gamma $.  Then whenever
that graph occurs as a subgraph of a bigger graph $\Gamma $, the
Feynman rules require that we also have a graph in which $\gamma $ is
replaced by its counterterm. Thus we must cancel the
subdivergences of a graph before asking how to cancel its overall
divergence.

\noindent 2.  After subtraction of subdivergences, the only
remaining divergence in a graph $\Gamma $ is its overall divergence.
Its strength is given by the overall degree of divergence $\Delta (\Gamma )$,
which is determined by simple power counting.

\noindent 3.  The overall divergence is cancelled by a new
contribution to the counterterm Lagrangian.

\noindent 4.  As already implied, counterterms to 1PI graphs are
implemented as counterterms in the Lagrangian.

\noindent 5.  Power counting for $\phi ^{4}$ in 4 space-time dimensions
shows, as we saw earlier at Eq.\ (\ref{eq:DegDivPhi4}), that the
only counterterms needed are those corresponding to terms in the
original Lagrangian.  That is, the theory is renormalizable to
all orders of perturbation theory.



\section{Renormalization Group}
\label{sec:RG}

We have seen in examples how one cancels the divergences in
Feynman graphs by adding divergent counterterms to the
Lagrangian.  Although the divergent parts of the counterterms are
determined by the requirement that we obtain finite renormalized
Green functions, the finite parts are not fixed.  So we have an
apparent freedom to make arbitrary changes in the results of
calculations by choosing the finite parts of the counterterms.  A
prescription for choosing the finite parts is called a
renormalization prescription or a renormalization scheme.

At first sight, the dependence of predictions on the choice of
renormalization scheme appears to remove the predictive power of
a theory, and hence to be undesirable physically.  In fact, as we
will see, a change of renormalization scheme can be completely
compensated by changes in the numerical values of the
finite renormalized parameters of the theory, which remains
invariant. The {\em renormalization group} is the formulation of
this invariance. A change of renormalization scheme is, in
essence, just a kind of non-linear change of the units for the
renormalized parameters. Another way of saying this is that it
amounts to a change of the splitting of the Lagrangian into basic
and counterterm parts Eq.\ (\ref{eq:LR}).

This suggests that, far from being a nuisance, renormalization
scheme dependence gives a powerful method for improving
perturbation calculations.  Consider the one loop calculation
Eq.\ (\ref{eq:1loop.R}).  At large $p^{2}$, the coefficient of $\lambda ^{2}$
has a logarithm of $p^{2}$, which ruins the accuracy of low-order
perturbation theory.  Note that higher order calculations get an
extra logarithm per loop --- see Eqs.\ (\ref{eq:2loop3})
and (\ref{eq:2loop3.Rbar}).  One change of renormalization
prescription is to change the value of the unit of mass $\mu $; then
we can try to get rid of the large logarithms by a choice of $\mu $.
This works in cases that all the momentum scales in a problem are
comparable.  In Sect.\ \ref{sec:OPE}, we will see some of the
techniques that can be used when there is more than one typical
scale in a problem.

When one changes $\mu $, one must compensate by changing the
coupling, so one comes to the concept of the {\em running
coupling}, or {\em effective coupling}, $\lambda (\mu )$.  Perturbative
calculations are valid if there are no large logarithms and if
the effective coupling is small.  One important result are the
renormalization group equations (RGEs), which are differential
equations that enable the running of the coupling, for example,
to be computed from elementary perturbative calculations.

In this section, we will explain the above concepts.

\subsection{Renormalization Schemes}

The calculation of the four point function in $\phi ^{4}$ theory can be
summarized as follows
\begin{eqnarray}
   G_{4} &=& -i\lambda
\label{eq:1loopA}
\\
   &&     - \frac {i\lambda ^{2}}{32\pi ^{2}}
             \left\{
                  \mbox{Pole}
                + \mbox{Complicated finite term}
                + \mbox{Counterterm}
             \right\}
          + O(\lambda ^{3}) .
\nonumber
\end{eqnarray}
Let us add and subtract a finite term $-ia\lambda ^{2}/(32\pi ^{2})$, with $a$
being an arbitrary finite number:
\begin{eqnarray}
   G_{4} &=& -i \left[\lambda  - \frac {a\lambda ^{2}}{32\pi ^{2}} \right]
\nonumber\\
      &&   - \frac {i\lambda ^{2}}{32\pi ^{2}}
              \left\{
                  \mbox{Pole}
                + \mbox{Complicated finite term}
                + \mbox{Counterterm}
                + a
              \right\}
\nonumber\\
      &&  + O(\lambda ^{3}) .
\label{eq:1loopB}
\end{eqnarray}
We have added an extra piece to the
counterterm and compensated by changing the value of the
coupling to
\begin{equation}
   \lambda ' = \lambda  - \frac {a\lambda ^{2}}{32\pi ^{2}} .
\label{eq:NewCoupling}
\end{equation}
We should, of course, change the coupling in the one-loop part of
Eq.\ (\ref{eq:1loopB}) from $\lambda $ to $\lambda '$.  At the level of
accuracy we are working to, $\lambda ^{2}$, the change will be of the same
as the higher order corrections that we have not yet calculated.

We make the following definitions:
\begin{quote}
   A {\em renormalization scheme} (or {\em prescription}) is a
   rule for deciding the finite part of a counterterm.
   Equivalently, it is a rule for defining the meaning of the
   renormalized parameters of a theory.
   \\
   {\em Renormalization group invariance} is the statement that
   when one changes the renormalization scheme, one keeps the
   same physics by suitable changes of $\lambda $, etc.
\end{quote}

When the external momenta are large, in our $\phi ^{4}$ example, we find
that
\begin{equation}
   G_{4} = -i\lambda
        - \frac {i\lambda ^{2}}{32\pi ^{2}} \left[
            \ln\frac {s}{\mu ^{2}} + \ln\frac {t}{\mu ^{2}} + \ln\frac {u}{\mu
^{2}} + \mbox{constant}
           \right]
        + O(\lambda ^{3}) .
\end{equation}
We can keep the coefficient of the one-loop term small by
choosing $\mu $, but only if the three invariants $s$, $t$, and $u$
are comparable in size.

\subsection{Particular renormalization schemes}

Among the infinity of possible renormalization schemes, only a
few are commonly used, and I list here some of the definitions

\paragraph{Physical Renormalization schemes}

A physical scheme is commonly used in QED.  There one defines the
renormalized mass of the electron to be exactly its physical
value.  The renormalized coupling $e$ is defined so that the long
range part of the electric field of an electron is the usual
$e/(4\pi r^{2})$ (in units with $\epsilon _{0}=1$).
In terms of Green functions, the definition of the mass is that
the pole in the electron propagator is at $p^{2} =  \BDpos m^{2}$, i.e., that the
self energy graphs are zero at $p^{2} =  \BDpos m^{2}$.
The definition of the coupling is that the 1PI $ee\gamma $ Green
function is equal to its lowest order value when the electrons
are on-shell and the momentum transfer to the photon is zero.

\paragraph{Momentum-space subtraction schemes}

Here, one defines the renormalized parameters to be equal to the
values of appropriate Green functions at some value of the
external momenta.  At the chosen value of external momentum, all
higher order corrections to the Green function are zero. A
physical renormalization scheme is one case of this.

\paragraph{Minimal Subtraction}

Minimal subtraction (MS) is defined given a particular
regularization scheme, normally dimensional regularization.
Counterterms are defined to be just the singular terms needed to
cancel the divergences.  With dimensional regularization, the
counterterms are just poles at $\epsilon =0$.  An extra mass scale, the
unit of mass $\mu $ needs to be introduced.

A common modification is the modified minimal subtraction scheme,
\MSbar, (with dimensional regularization).  Here one redefines
$\mu $ by a particular factor to cancel certain numerical terms that
typically arise in a calculation using dimensional
regularization.

Minimal subtraction can also be defined with other regulators.
For example, with a lattice spacing $a$, one could define the
counterterms in each order of perturbation theory to be
polynomials in $1/a$ and $\ln(a\mu )$, with no constant term.

Minimal subtraction has the advantage that the counterterms only
contain the divergences.  It has the disadvantage of being purely
perturbative in formulation and of depending on the choice of
regulator.

\subsection{Renormalization Group}

The discussion in this and the next few sections will be written
out for the $\phi ^{4}$ theory, with dimensional regularization as the
ultra-violet regulator.  It will be evident that the whole
treatment can be generalized to other cases with only notational
changes.

In the bare Lagrangian Eq.\ (\ref{eq:L0}) there are exactly two
parameters, $m_{0}$ and $\lambda _{0}$.  Renormalization is done by allowing
the bare parameters to be adjusted suitably as the ultra-violet
cutoff is removed, and by defining a renormalized field
$\phi =\phi _{0}/\sqrt Z$, with the ``wave-function renormalization'' factor
$Z$
also being allowed cutoff dependence.

Suppose now that we have two renormalization schemes, which we
will label as 1 and 2.  The renormalized coupling and mass in
each of these schemes we write as $\lambda _{1}$, $m_{1}$, $\lambda _{2}$ and
$m_{2}$.  The
wave function renormalizations are $Z_{1}$ and $Z_{2}$

Renormalized Green functions, i.e., time-ordered
vacuum-expectation values of the renormalized fields, are readily
related in the two schemes:
\begin{eqnarray}
   G_{N}^{(1)}(p, \lambda _{1},m_{1}) &=& Z_{1}^{-N/2} G_{N}^{(0)}(p, \lambda
_{0},m_{0})
\nonumber\\
   &=& \left[ \frac {Z_{1}(\lambda _{1},m_{1};\epsilon )}{Z_{2}(\lambda
_{2},m_{2};\epsilon )} \right]^{-N/2}
       G_{N}^{(2)}(p, \lambda _{2},m_{2}) .
\label{eq:RSChange}
\end{eqnarray}
Here $G_{N}^{(1)}$ and $G_{N}^{(2)}$ denote the renormalized $N$-point Green
functions, while $G_{N}^{(0)}$ denotes the bare Green function.

Initially, we work with $\epsilon =0$ so that all parts of this formula
make sense, and then we will take the limit as $\epsilon \to 0$, so that
only the renormalized Green functions can be used.

The three different Green functions in Eq.\ (\ref{eq:RSChange})
are related by changes of variables:
\begin{equation}
   (\lambda _{1},m_{1}) \mbox{~to~} (\lambda _{0},m_{0}) \mbox{~to~} (\lambda
_{2},m_{2}) .
\end{equation}
Now the renormalized Green functions are finite functions of their
renormalized parameters as $\epsilon =0$.  In order for this to be
consistent with the relation between the left- and the right-hand
sides of Eq.\ (\ref{eq:RSChange}), $\lambda _{1}$ and $m_{1}$ must be finite
functions of $\lambda _{2}$ and $m_{2}$ as $\epsilon \to 0$.  Furthermore
$Z_{1}/Z_{2}$ must be
a finite function of $\lambda _{2}$ and $m_{2}$ (and hence a finite function
of $\lambda _{1}$ and $m_{1}$).

Thus we can write
\begin{equation}
   G_{N}^{(1)}(p, \lambda _{1},m_{1}) = z_{12}^{N/2} G_{N}^{(2)}(p, \lambda
_{2},m_{2}) ,
\label{eq:RG}
\end{equation}
an equation in which all quantities are finite at $\epsilon =0$, and
$z_{12}=Z_{2}/Z_{1}$ is a calculable function of the renormalized
parameters.  This equation is the most fundamental expression of
renormalization-group invariance.  It says exactly that a change
of renormalization scheme can be compensated by suitable changes
in the renormalized parameters of the theory ($\lambda $ and $m$)
together with a rescaling of the renormalized field by a factor:
\begin{equation}
   \phi _{1}(x) = \sqrt {z_{12}} \phi _{2}(x) .
\end{equation}

\subsubsection{Invariance of Physical Mass and of $S$-matrix}

The singularities of the Green functions as a function of
external momenta must be at the same positions in the two
schemes, by Eq.\ (\ref{eq:RG}).  In particular the physical
mass of the particle is the same.

To show invariance of the $S$-matrix, we use the LSZ reduction
formula.  We need the residues of the propagator poles, which we
define by:
\begin{equation}
   G_{2}^{(i)} \sim \frac {i c^{(i)}}{ \BDpos p^{2} - m_{\rm ph}^{2}}
   \mbox{~as $ \BDpos p^{2}\to m_{\rm ph}^{2}$} .
\end{equation}
Eq.\ (\ref{eq:RG}) then implies that the residues are related by
\begin{equation}
   c^{(1)} = z_{12}c^{(2)} .
\end{equation}

The S-matrix in scheme 1 is
\begin{equation}
   S_{2\to N}^{(1)} =
   \lim_{\rm on-shell}
   \frac {[c^{(1)}]^{(N+2)/2} \, G_{N+2}^{(1)}}{\prod 2+N {\rm~propagators}} .
\label{eq:RGS}
\end{equation}
Each factor on the right of this equation is related to the
corresponding quantity in scheme 2 by multiplication by a power
of $z_{12}$, and the overall power of $z_{12}$ is zero, so that the
$S$-matrix is the same in the two schemes.

We conclude then that the $S$-matrix (and hence cross sections)
are renor\-mal\-iza\-tion-group invariant.

\subsection{Running Coupling and Mass}

One possibility for a change of renormalization scheme is to stay
within the MS scheme, but to change the value of the unit of mass
$\mu $.  (Exactly similar methods can be employed in any other
scheme, of course.)  To keep the physics fixed, our discussion in
the previous section implies that we must adjust $\lambda $ and $m$ as
$\mu $ varies.  That is, we must have $\lambda =\lambda (\mu )$ and $m=m(\mu
)$, the
running (or effective) coupling and mass.

Eq.\ (\ref{eq:RG}) tells us the form of the relation between the
Green functions with different values of $\mu $:
\begin{equation}
   G_{N}(p, \lambda (\mu '), m(\mu '), \mu ')
   =
   z(\mu '/\mu , \lambda (\mu ))^{N} \,
   G_{N}(p, \lambda (\mu ), m(\mu ), \mu ) ,
\label{eq:RGE0}
\end{equation}
where $z(\mu '/\mu , \lambda (\mu )) = \lim_{\epsilon \to 0} \sqrt {\strut
Z(\lambda (\mu ),\epsilon ) / Z(\lambda (\mu '),\epsilon )}$.
By differentiating with respect to $\mu $, we obtain the standard
renormalization-group equation:
\begin{equation}
   \mu \frac {d}{d\mu }G_{N} = - \frac {N}{2} \gamma (\lambda (\mu )) G_{N} ,
\label{eq:RGE}
\end{equation}
where
\begin{equation}
   \frac {1}{2}\gamma  = - \mu '\frac {\partial }{\partial \mu '}z(\mu '/\mu
,\lambda (\mu ))\Bigg|_{\mu '=\mu }
\label{eq:gammaDef}
\end{equation}
is called the ``anomalous dimension'' of the field $\phi $.  The
factor $1/2$ in its definition is one standard convention.  We will
see the rationale for the name later.
The derivative in Eq.\ (\ref{eq:RGE}) is a total derivative: it
acts on the argument of the running coupling and mass as well as
on the explicit $\mu $ argument of the Green function.

Given the differential renormalization-group equation Eq.\
(\ref{eq:RGE}), we can reconstruct Eq.\ (\ref{eq:RGE0})
\begin{equation}
   G_{N}(p, \lambda (\mu '), m(\mu '), \mu ')
   =
   G_{N}(p, \lambda (\mu ), m(\mu ), \mu )
   \exp\left[
          - \frac {N}{2} \int _{\mu }^{\mu '} \frac {d\hat\mu }{\hat\mu }
\gamma (\lambda (\hat\mu ))
       \right] .
\label{eq:RGESolution}
\end{equation}

\subsection{Computation of Running Coupling and Mass}

To make Eq.\ (\ref{eq:RGESolution}) an effective way of
performing calculations, we must have a convenient way of
obtaining the effective coupling $\lambda (\mu )$ and the effective mass
$m(\mu )$.
First we observe that the bare coupling $\lambda _{0}$ is
renormalization-group invariant.  Dimensional analysis and the
knowledge that counterterms are polynomial in mass when minimal
subtraction is used implies that the bare coupling has the form
$\lambda _{0}=\mu ^{\epsilon }\bar\lambda (\lambda ,\epsilon )$.  Hence
renormalization-group invariance of
$\lambda _{0}$ gives
\begin{equation}
   0 = \mu \frac {d\lambda _{0}}{d\mu }
     = \mu ^{\epsilon } \left[
             \epsilon \bar\lambda  + \mu \frac {d\lambda (\mu )}{d\mu } \frac
{\partial \bar\lambda }{\partial \lambda }
          \right] .
\end{equation}
Hence
\begin{equation}
   \mu \frac {d\lambda (\mu )}{d\mu } = - \frac {\epsilon \lambda _{0}(\lambda
,\epsilon )}{\partial \lambda _{0}/\partial \lambda } .
\label{eq:RGBeta}
\end{equation}
The right-hand side is usually denoted by $\beta (\lambda ,\epsilon )$, and it
must
be finite as $\epsilon \to 0$, so that the renormalization group
variation of the coupling does not involve any divergences.  The
divergences on the right-hand side of Eq.\ (\ref{eq:RGBeta})
cancel, and this must occur order-by-order in perturbation
theory.

When we renormalize the theory and calculate the bare coupling to
some order in $\lambda $, we can obtain the renormalization-group
coefficient $\beta $ to the same order.  Then we can use this
approximation to $\beta $ to compute, to some accuracy, the running
coupling according to:
\begin{equation}
   \int _{\lambda (\mu )}^{\lambda (\mu ')} \frac {d\lambda }{\beta (\lambda
,0)} = \ln\frac {\mu '}{\mu } ,
\end{equation}
which is obtained by integrating the differential equation Eq.\
(\ref{eq:RGBeta}).

We may similarly obtain the running mass and its renormalization
group coefficient from
\begin{equation}
   \mu \frac {dm^{2}(\mu )}{d\mu } = \gamma _{m}(\lambda ) m^{2}
\end{equation}
with
\begin{equation}
   \gamma _{m}(\lambda ) = -\beta (\lambda ,\epsilon ) \frac {\partial  \ln
m_{0}^{2}/m^{2}}{\partial \lambda } .
\end{equation}
The lack of $\epsilon $ dependence for $\gamma _{m}$ is due to the specifics of
minimal subtraction.

We may also derive a formula for the anomalous dimension:
\begin{equation}
   \gamma (\lambda ) = \beta (\lambda ,\epsilon ) \frac {\partial  \ln
Z(\lambda ,\epsilon )}{\partial \lambda } .
\end{equation}

\subsection{Renormalization-Group Coefficients in $\phi ^{4}$ theory}

Up to two-loop order, the bare coupling is\footnote{Terms up to
   five-loop order are now known\protect.\cite{phi4RG}}
\begin{equation}
   \lambda _{0} = \mu ^{\epsilon } \left[
               \lambda  + \frac {3}{\epsilon } \frac {\lambda ^{2}}{16\pi ^{2}}
               +\frac {\lambda ^{3}}{(16\pi ^{2})^{2}} \left( \frac
{9}{\epsilon ^{2}} - \frac {17}{6\epsilon } \right)
               + O(\lambda ^{4})
           \right] .
\end{equation}
It follows that the $\beta $ function is
\begin{eqnarray}
 \beta  &=& \frac {-\epsilon  \left[ \lambda  + \frac {3}{\epsilon } \frac
{\lambda ^{2}}{16\pi ^{2}} + \frac {\lambda ^{3}}{(16\pi ^{2})^{2}} \left(
\frac {9}{\epsilon ^{2}} - \frac {17}{6\epsilon } \right) + O(\lambda ^{4})
\right]}{1 + \frac {6}{\epsilon } \frac {\lambda }{16\pi ^{2}} + \frac {\lambda
^{2}}{(16\pi ^{2})^{2}} \left( \frac {27}{\epsilon ^{2}} - \frac {17}{2\epsilon
} \right) + O(\lambda ^{4})}
\nonumber\\
   &=& -\epsilon \lambda  + \frac {3 \lambda ^{2}}{16\pi ^{2}} - \frac {17}{3}
\frac {\lambda ^{3}}{(16\pi ^{2})^{2}} + O(\lambda ^{4}) .
\end{eqnarray}

\subsection{Solution for Running Coupling}

Given the renormalization-group equation
\begin{equation}
   \mu \frac {d\lambda }{d\mu } = \beta (\lambda (\mu )) ,
\end{equation}
and a finite-order approximation to $\beta $, we may get an
approximation for the running of the coupling.  We can estimate
the errors in the solution, when $\lambda $ is small, from knowing the
order of magnitude of uncalculated higher-order terms in $\beta $.

\begin{figure}
    \begin{center}
    \leavevmode
        \rlap{\raisebox{0.5in}{\hspace*{0.0in}$\beta (\lambda )$}}
        \rlap{\raisebox{-0.4in}{\hspace*{1.25in}$\lambda $}}
        \epscenterbox{1.2in}{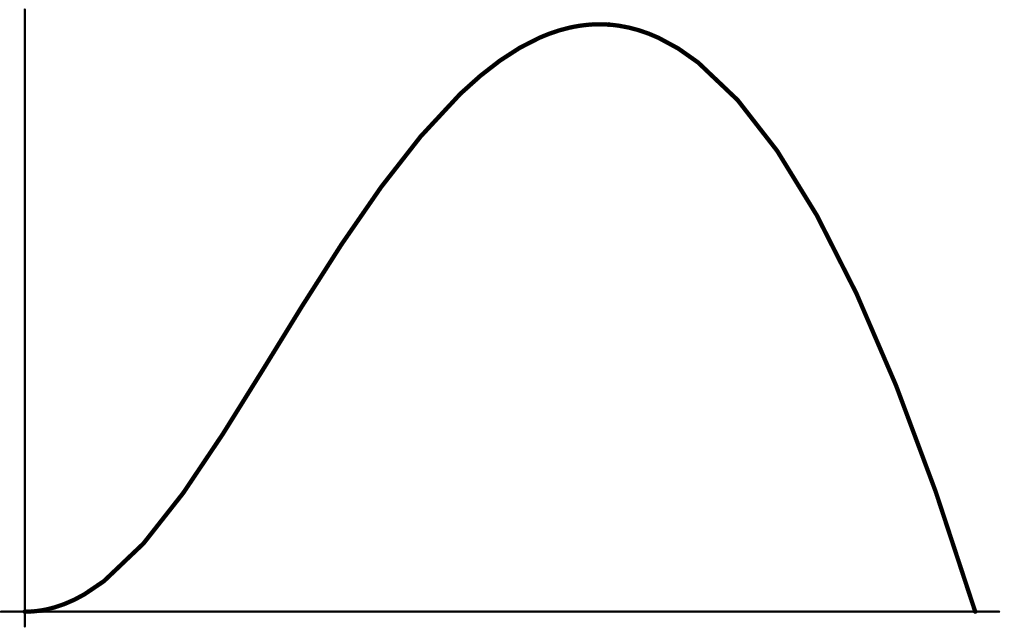}
    \hspace*{0.8in}
        \rlap{\raisebox{0.5in}{\hspace*{0.0in}$\lambda (\mu )$}}
        \rlap{\raisebox{-0.4in}{\hspace*{1.25in}$\ln \mu $}}
        \epscenterbox{1.2in}{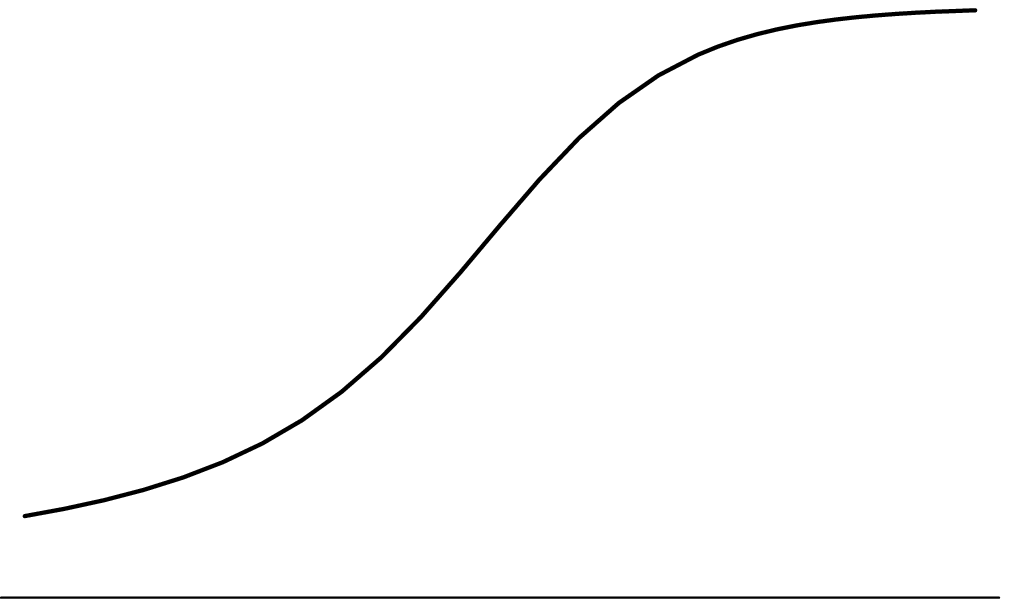}
   \end{center}
   \caption{Running coupling in $\phi ^{4}$ on the assumption of an
            ultra-violet fixed point. }
\label{fig:Run.phi4}
\end{figure}

We remove the regulator, and observe that $\beta $ is proportional to
$\lambda ^{2}$, with a positive coefficient, for small $\lambda $.  For large
$\lambda $,
we are out of the range of accuracy of finite-order calculations.
We show in Fig.\ \ref{fig:Run.phi4} what happens to the running
of the coupling if $\beta $ has a zero at some non-zero $\lambda =\lambda
^{*}$.  For
small $\mu $, $\lambda (\mu )$ is small and approximately proportional to
$1/\ln(1/\mu )$; this is a firm prediction in a region where
perturbation theory is accurate. We call $\lambda =0$ an infra-red fixed
point. At large $\mu $, $\lambda (\mu )\to \lambda ^{*}$.  We therefore call
$\lambda ^{*}$ an
``ultra-violet fixed point''.  If $\beta $ did not have a zero, then
the coupling would increase without limit as $\mu \to \infty $.

\subsection{Asymptotic Freedom}

Suppose we have theory in which $\beta $ is {\em negative} for small
coupling.  An example is QCD, where we would use the notation
$\alpha _{s}$ instead of $\lambda $.  Then zero coupling is an ultra-violet
fixed
point; this is the defining property of an ``asymptotically
free'' theory.
As shown in Fig.\ \ref{fig:Run.phi4}, the coupling goes to zero
as $\mu \to \infty $.

In an asymptotically free theory we may use the renormalization
group to allow accurate calculations of ultra-violet dominated
quantities, as we will see in a moment.

\begin{figure}
    \begin{center}
    \leavevmode
       \rlap{\raisebox{0.5in}{\hspace*{0.0in}$\beta (\alpha _{s})$}}
       \rlap{\raisebox{0.35in}{\hspace*{1.25in}$\alpha _{s}$}}
       \epscenterbox{1.2in}{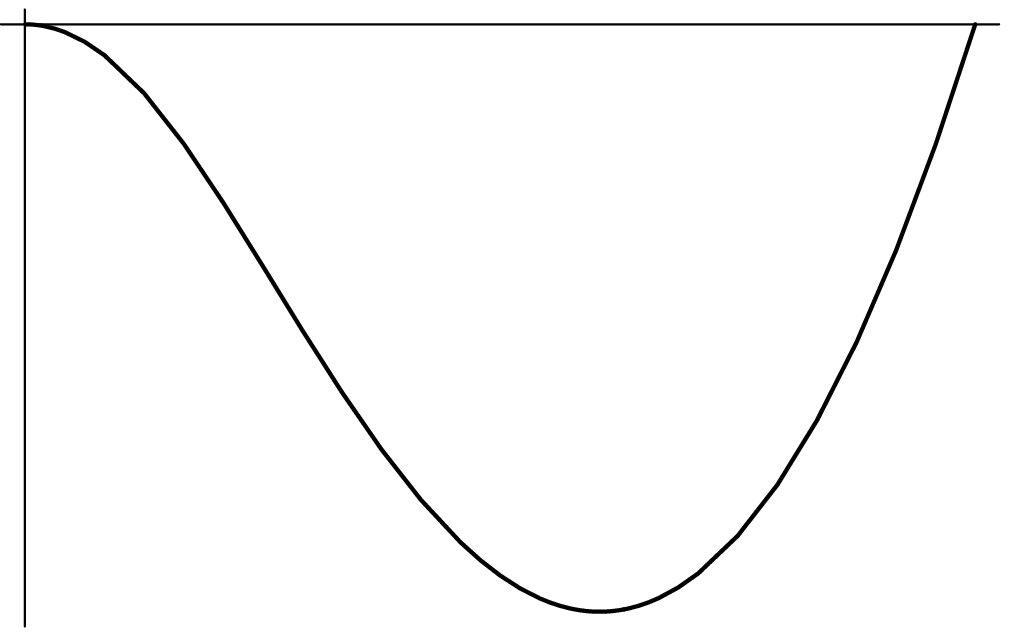}
    \hspace*{0.8in}
       \rlap{\raisebox{0.5in}{\hspace*{0.0in}$\alpha _{s}(\mu )$}}
       \rlap{\raisebox{-0.4in}{\hspace*{1.25in}$\ln \mu $}}
       \epscenterbox{1.2in}{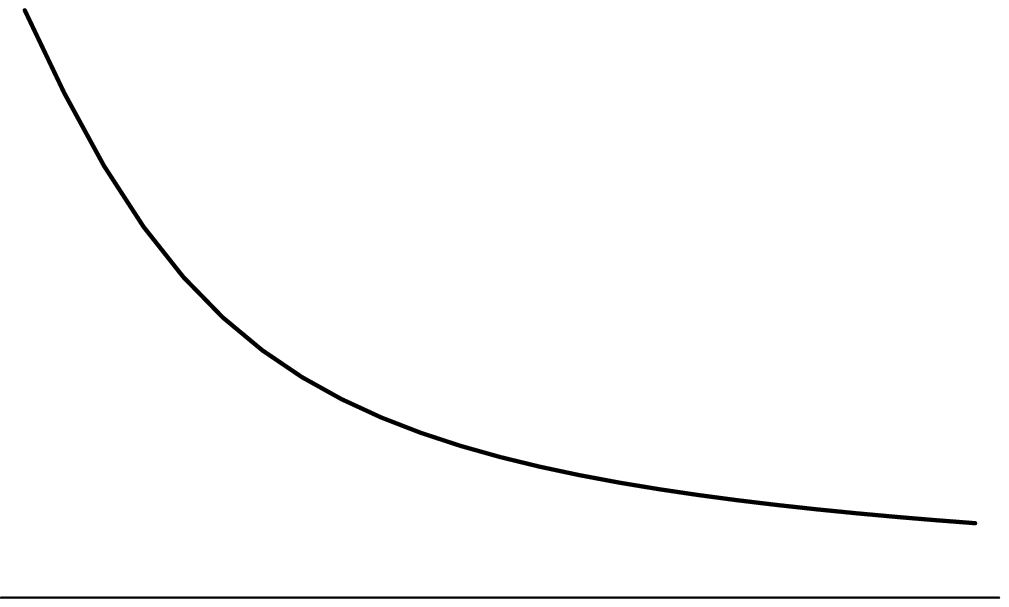}
   \end{center}
   \caption{Running coupling in an asymptotically free theory.}
\label{fig:Run.QCD}
\end{figure}

\subsection{Large Momentum Behavior}

Suppose we wish to calculate the large momentum behavior of a
Green function, for example the two point function $G_{2}(p^{2})$ as
$|p^{2}|\to \infty $.  We use the solution of the renormalization group
equation (\ref{eq:RGESolution}) to replace a fixed $\mu $ by
$\sqrt {|p^{2}|}$:
\begin{eqnarray}
\lefteqn{
   G_{2}(p^{2}, \lambda (\mu ), m(\mu ), \mu )
}~~
\nonumber\\
   &=&
   G_{2}\left(  \BDpos p^{2}, \lambda \left(\sqrt {|p^{2}|}\right),
            m\left(\sqrt {|p^{2}|}\right), \sqrt {|p^{2}|}
      \right)
   \, e^{\int _{\mu }^{\sqrt {|p^{2}|}} \frac {d\mu '}{\mu '} \gamma
\left(\lambda (\mu ')\right)}
\nonumber\\
    &\simeq&
    G_{2}\left( p^{2}, \lambda \left(\sqrt {|p^{2}|}\right), 0, \sqrt {|p^{2}|}
\right)
    \, e^{\int _{\mu }^{\sqrt {|p^{2}|}} \frac {d\mu '}{\mu '} \gamma
\left(\lambda (\mu ')\right)}
\nonumber\\
   &\simeq&
    \frac {1}{|p^{2}|} G_{2}\left( p^{2}/|p^{2}|, \lambda \left(\sqrt
{|p^{2}|}\right), 0, 1 \right)
    \, e^{\int _{\mu }^{\sqrt {|p^{2}|}} \frac {d\mu '}{\mu '} \gamma
\left(\lambda (\mu ')\right)} ,
\label{eq:RG2}
\end{eqnarray}
where in the second line we have neglected the mass, as is
sensible at large $p^{2}/m^{2}$ and in the last line we simply used
dimensional analysis.

If we have a theory with a UV fixed point $\lambda ^{*}\not=0$, then the
large $p$ behavior is
\begin{equation}
   G_{2}(p^{2}, \lambda (\mu ), m(\mu ), \mu )
   \simeq \mbox{constant} \times  (p^{2})^{-1+\frac {1}{2}\gamma (\lambda ^{*})
}.
\end{equation}
The power law would be obtained from the na\"{\i}vest
considerations of dimensional analysis if the dimension of the
field were increased by $\gamma /2$; this accounts for the terminology
``anomalous dimension''.

If we were working in an asymptotically free theory, then
$\lambda (\sqrt {|p^{2}|}) \to  0$ as $p^{2}\to \infty $, and we would be able
to use weak
coupling perturbation theory to estimate the right-hand side of
Eq.\ (\ref{eq:RG2}), up to an overall factor.  The power law
would be $1/p^{2}$, but modified by logarithmic corrections from the
integral over the anomalous dimension function.

\subsection{Coupling in asymptotically free theory}

I will conclude by indicating a few properties of an
asymptotically free theory, using QCD as an example.
There the running coupling obeys
\begin{equation}
   \mu \frac {d}{d\mu }\alpha _{s}(\mu ) = - \frac {\beta _{0}}{2\pi } \alpha
_{s}^{2} - \frac {\beta _{1}}{4\pi ^{2}} \alpha _{s}^{3} - O(\alpha _{s}^{4}) ,
\end{equation}
where in fact the first three coefficients on the right are
known\cite{PDG}.  A solution of this equation has one constant of
integration.  It is common to define it by writing the solution
as
\begin{equation}
   \alpha _{s}(\mu ) = \frac {4\pi }{\beta _{0} \ln \mu ^{2}/\Lambda ^{2}}
           - \frac {8\pi  \beta _{1} \ln (\ln \mu ^{2}/\Lambda ^{2})}{\beta
_{0}^{3} \ln^{2}(\mu ^{2}/\Lambda ^{2})}
           + O\left( \frac {\ln^{2}(\ln \mu )}{\ln^{3} \mu } \right) .
\end{equation}
The boundary condition determining the numerical value of the
constant $\Lambda $ is that there is no $1/\ln^{2}(\mu ^{2}/\Lambda ^{2})$
term.  One
parameter $\Lambda $ determines the whole function $\alpha _{s}(\mu )$.

As stated earlier, one can then use the renormalization group to
allow the use of perturbation theory to compute large-momentum
behavior.  The condition for the applicability of perturbation
theory for computing behavior at some scale $Q$ is that $\alpha _{s}/\pi $ is
sufficiently less than unity.

\subsection{Bare coupling in asymptotically free theory}

One nice application of the renormalization group is to compute
the true bare coupling.  This means that one computes how the
bare coupling behaves as the regulator is removed with $\alpha _{s}$ being
held fixed. Recall that the bare coupling is expressed as a
perturbation series in the renormalized coupling $\alpha _{s}$, so fixed
order perturbation theory does not directly predict the true
behavior of the bare coupling. (Perturbation theory is {\em a
priori} about the limit $\alpha _{s}\to 0$ with the regulator fixed.)

Let us suppose we use a lattice regulator.  (We can do a similar
calculation with dimensional regularization, but the result is
less intuitive.)  The momentum scale relevant to the bare
coupling is the inverse of the lattice spacing $a$. We start with
the bare coupling $\alpha _{s(0)}(\alpha _{s},a\mu )$ at the values of $\alpha
_{s}$ and $\mu $
that we are interested in.  Then we use a renormalization-group
transformation to to reexpress it in terms of the running
coupling at $\mu =1/a$.  Asymptotic freedom then enables us to get a
useful result from low order perturbation theory.

We can write the bare coupling in terms of the \MSbar{} coupling
in the form
\begin{equation}
   \alpha _{s(0)} = \alpha _{s}
        + \frac {\beta _{0}\alpha _{s}^{2}}{2\pi } \left[ \ln(a\mu ) + C_{0}
\right]
        + O(\alpha _{s}^{3}) ,
\end{equation}
where the constant $C_{0}$ is known.  Renormalization-group
invariance of the bare coupling enables us to write this as
\begin{eqnarray}
   \alpha _{s(0)}(\alpha _{s}(\mu ), a\mu ) &=& \alpha _{s(0)}(\alpha
_{s}(1/a), 1)
\nonumber\\
   &=&  \alpha _{s}(1/a)
        + \frac {C_{0} \beta _{0}\alpha _{s}^{2}(1/a)}{2\pi }
        + O(\alpha _{s}^{3}(1/a))
\nonumber\\
   &=& \frac {4\pi }{\beta _{0} \ln 1/a^{2}\Lambda ^{2}}
           - \frac {8\pi  \beta _{1} \ln (\ln 1/a^{2}\Lambda ^{2})}{\beta
_{0}^{3} \ln^{2}(1/a^{2}\Lambda ^{2})}
           + \frac {8\pi  C_{0}}{\beta _{0} \ln^{2}(1/a^{2}\Lambda ^{2})}
\nonumber\\
           &&~~~+ \dots .
\end{eqnarray}
An important result is that the omitted terms can be omitted
without affecting the $a\to 0$ limit of the renormalized theory.
Hence it is sufficient to know the $\beta $ function to two-loop
order, and the relation between the renormalization of the
lattice theory and \MSbar{} renormalization at one-loop order.

For reference, note that
\begin{equation}
   \beta _{0} = 11 - \frac {2}{3}n_{f};~~~\beta _{1} = 51 - \frac {19}{3}n_{f}
,
\end{equation}
where $n_{f}$ is the number of quark flavors.  A reasonable fit to
the measured coupling, with $n_{f}=5$, is obtained with
$\Lambda _{\MSbar}=0.5{\rm\,GeV}$.  This implies that $\alpha _{s}(m_{Z})
\simeq
0.117\pm 0.005$.

\subsection{Summary}

\noindent 1.  Renormalizable theories are renormalization-group
invariant. That is, the physics they predict is unchanged under a
change of the renormalization scheme, provided suitable changes
in the values of the renormalized coupling and mass are made.

\noindent 2.  This leads to the concepts of the running coupling
and mass.

\noindent 3.  We can use this information to compute and analyze
mo\-men\-tum-de\-pen\-dence.  One particular case is the
computation of the true properties of the bare parameters as the
regulator is removed.

\noindent 4.  A very important case is that of asymptotically free
theories, where the running coupling goes to zero at infinite
momentum.



\section{Operator Product Expansion (OPE)}
\label{sec:OPE}

The theory of the strong interactions of hadrons, QCD, is
asymptotically free theory. This implies that we can use the
renormalization group to calculate a Green function, for example,
if all its external momenta get large.  However, normal physical
processes never have all their momenta large, since scattering
experiments have to keep their beams and targets on-shell.  The
operator product expansion is the simplest of a number of
theorems on asymptotic behavior in field theory that enable
useful predictions to be made in QCD.

\begin{figure}
    \begin{center}
        \leavevmode
        \epscenterbox{1.3in}{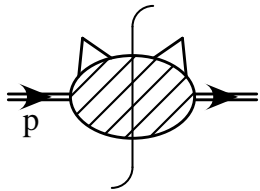}
    \end{center}
    \caption{Hadronic part of deep-inelastic scattering.}
    \label{fig:Wmunu}
\end{figure}

A simple case is deep-inelastic scattering of an electron off a
hadron: $e+p\to e'+X$.  We work in the
good approximation of single photon exchange, and let $p^{\mu }$ be the
target hadron's momentum and let $q^{\mu }$ be the exchanged photon's
momentum.  We choose to sum over all possible hadronic
final-states $X$ given $p^{\mu }$ and $q^{\mu }$.  Then the strong
interaction part of the cross section is obtained from the
Green function depicted in Fig.\ \ref{fig:Wmunu}.  The
deep-inelastic limit is $\BDneg q^{2}\to \infty $ with $x\equiv -q^{2}/2p\cdot q$
fixed.

\subsection{Simplest case}

To see the relevant field-theoretic principles at work in their
simplest form we consider the following Green function in $\phi ^{4}$
theory:
\begin{equation}
   W(p,q)
   =\int d^{4}x \, d^{4}y_{1} \, d^{4}y_{2} \,
   e^{\BDneg iq\cdot x \BDminus ip\cdot y_{1} \BDplus ip\cdot y_{2}}
    \langle 0| T j(x) j(0) \phi (y_{1}) \phi (y_{2}) |0\rangle  ,
\end{equation}
where $j=\half\phi ^{2}$ is used instead of the electromagnetic current, and
the two $\phi $ factors are used as interpolating fields for the
target.  We have set the positions of one of the four fields
equal to zero, to avoid having to carry around a delta-function
for momentum conservation.

We will choose to take a limit in which we scale $q$ to infinity:
$q^{\mu }=\lambda q_{0}^{\mu }$, with $\lambda \to \infty $ at fixed
$q_{0}^{\mu }$ and fixed $p^{\mu }$.  This is not
exactly the deep-inelastic limit, but a dispersion relation can
be used to relate the results of the OPE we derive in this limit
to moments of deep-inelastic structure functions.\footnote{
    The limit we take can be characterized as ``Euclidean'', in
    contrast to the Bjorken limit of deep-inelastic scattering,
    which is only possible for Minkowskian momenta.
    The Bjorken limit has both $q^{2}$ and $p\cdot q$ scaling as $\lambda
^{2}$,
    with $\lambda \to \infty $, whereas for Euclidean vectors one must have
$p\cdot q$
    scaling only as $\lambda $, if $q^{2}\propto \lambda ^{2}$.  Asymptotic
problems in
    Minkowski space are treated in Sterman's lectures, where they
    will be seen to be substantially more complex.
}

The OPE has the form
\begin{eqnarray}
   \epscenterbox{0.8in}{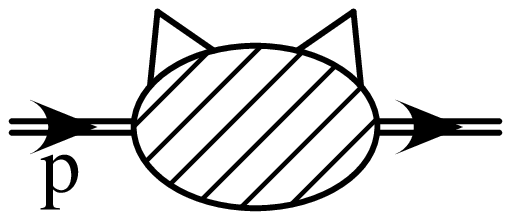}
   &=& \sum_{i} C_{i}(q) \int  d^{4}y_{1} \, d^{4}y_{2} \, 
       e^{ \BDneg ip\cdot y_{1} \BDplus ip\cdot y_{2}}
    \langle 0| T {\cal O}_{i}0) \phi (y_{1}) \phi (y_{2}) |0\rangle
\nonumber\\
   &=& \sum_{i} C_{i}(q)
   ~~\epscenterbox{0.8in}{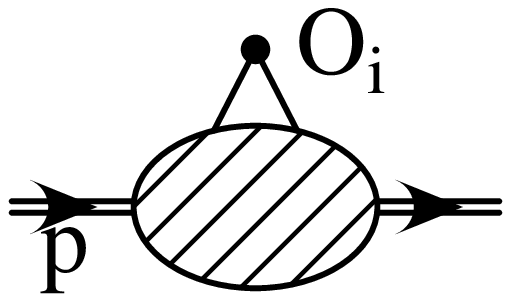} ,
\label{eq:OPE}
\end{eqnarray}
where the sum runs over all local operators (${\cal O}_{i}= \phi ^{2},\
(\partial \phi )^{2},\ \phi ^{4}$ etc).  The predictive power of the OPE will
come
from the fact that the coefficient functions $C_{i}(q)$ have greater
power law suppressions as the dimension of ${\cal O}_{i}$ gets
larger.  Only a small number of terms is relevant, and
the renormalization group plus finite-order perturbation theory
may be used to compute the coefficients to a useful
approximation. The product of two fields $\phi (y_{1}) \phi (y_{2})$ may be
replaced by a product of any number of fields of fixed momenta;
these would define the target system in deep-inelastic
scattering.

\subsubsection{Lowest order}

We can get basic intuition about how the OPE arises by examining
tree graphs.  For example, the lowest order graphs give:
\begin{eqnarray}
\lefteqn{
   \epscenterbox{1.0in}{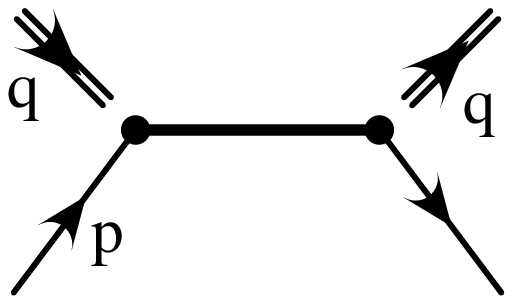}
   ~+~ \epscenterbox{1.0in}{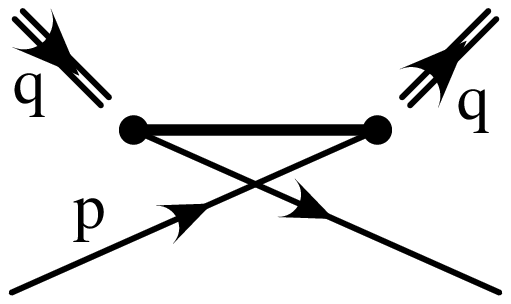}
}
\nonumber\\
   &=& \left( \frac {i}{ \BDpos p^{2}-m^{2}} \right)^{2}
       \left[
         \frac {i}{\BDpos (q+p)^{2}-m^{2}} + \frac {i}{(q-p)^{2}-m^{2}}
       \right]
\nonumber\\
   &=& \left( \frac {i}{ \BDpos p^{2}-m^{2}} \right)^{2}
       \left[
         \frac {2i}{\BDpos q^{2}}
         + \frac {2i}{(q^{2})^{2}} \left( m^{2} \BDminus p^{2} \BDplus \frac {4p\cdot
q^{2}}{q^{2}} \right)
         +O \left( \frac {1}{q^{6}} \right)
       \right]
\nonumber\\
 &=& \frac {2i}{\BDpos q^{2}}
     ~\rlap{\raisebox{0.15in}{\hspace*{0.35in}$\frac {1}{2}\phi ^{2}$}}
      \epscenterbox{0.5in}{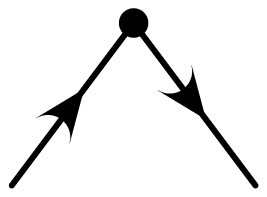}
     ~~+~~ \frac {2i}{(q^{2})^{2}}
     ~\rlap{\raisebox{0.15in}{\hspace*{0.35in}$\frac {1}{2}(m^{2}\phi
^{2} \BDneg (\partial \phi )^{2})$}}
      \epscenterbox{0.5in}{sf-1-op.eps}
\nonumber\\
 & & +~~ \frac {8iq^{\mu }q^{\nu }}{(\BDpos q^{2})^{3}}
      ~\rlap{\raisebox{0.15in}{\hspace*{0.35in}$\frac {1}{2}\partial _{\mu
}\phi \partial _{\nu }\phi $}}
      \epscenterbox{0.5in}{sf-1-op.eps}
     ~~~~~~~~~~+~ \dots .
\label{eq:tree2}
\end{eqnarray}
Here we have expanded in powers of the small variables, $m/q$,
and $p/q$.  In applications, we typically only keep the first
term in the series, the term of order $1/q^{2}$ in Eq.\
(\ref{eq:tree2}).  In the diagrams, I have used thick lines to
denote the lines that are forced to carry large momentum.

There is a pattern here, which we can represent in the form
\begin{equation}
   \epscenterbox{1.0in}{dis-sm.eps}
   ~~~=~~ \epscenterbox{1.0in}{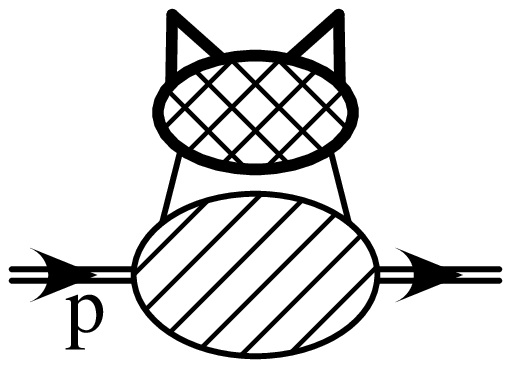}
\label{eq:OPEGraph}
\end{equation}
On the left-hand side we have a general graph for the left-hand
side of Eq.\ (\ref{eq:OPE}).  On the right, we have split it into
two factors: The top factor, marked by thick lines, carries large
momentum, of order $Q$, and is called the ``hard part''.  The
bottom factor carries momenta of low virtuality, and it is called
the ``soft part''. By expanding the hard part in powers of its
soft external momenta, we obtain terms in the Wilson coefficients
$C_{i}$.  Then the soft part turns into the matrix elements of the
operators ${\cal O}_{i}$ in Eq.\ (\ref{eq:OPE}).

All the $q$ dependence is in the coefficients $C_{i}(q)$, and all the
$p$ and $m$ dependence is in the Green functions or matrix
elements of the operators ${\cal O}_{i}$.
Just as with the ultra-violet divergences, the power of $q$ in
$C_{i}(q)$ may be determined by dimensional analysis, as follows.
Suppose $C_{i}(q) \propto  1/q^{N_{i}}$.
\begin{equation}
   -N_{i} = 2 \, {\rm dim}(j) - 4 - {\rm dim}({\cal O}_{i}) ,
\end{equation}
where the ``4'' comes from the $x$ integral for the Fourier
transform.  A particular consequence of this is that the leading
power of $q$ is in the term with the lowest dimension operator.
This strongly restricts the number of terms

\subsubsection{Is the pattern general?}

We may test the pattern just conjectured by examining other
graphs.  Consider the case of four external low momentum
lines. To lowest order we find:
\begin{eqnarray}
   \epscenterbox{0.9in}{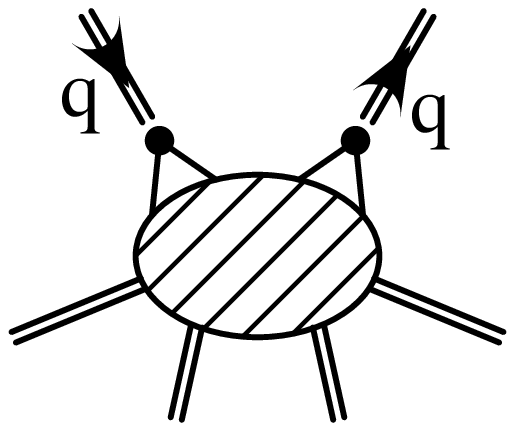}
   &=& \epscenterbox{1.0in}{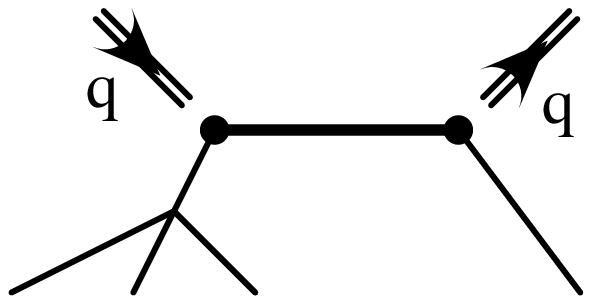}
       ~+~ \epscenterbox{1.0in}{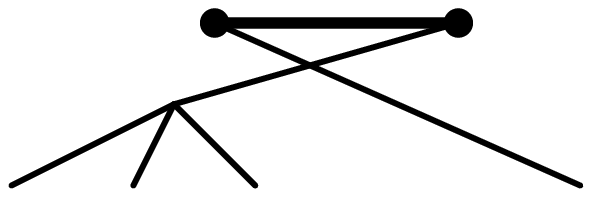}
       ~+~ \mbox{6 similar}
\nonumber\\[0.03in]
   & & +~ \epscenterbox{1.0in}{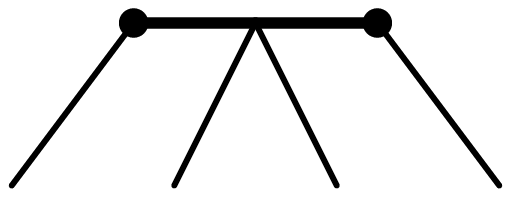} ~+~ \mbox{11 similar}
\nonumber\\[0.1in]
   &=& \frac {2i}{q^{2}} \left[
             ~\rlap{\raisebox{0.20in}{\hspace*{0.65in}$\frac {1}{2}\phi ^{2}$}}
             \epscenterbox{1.0in}{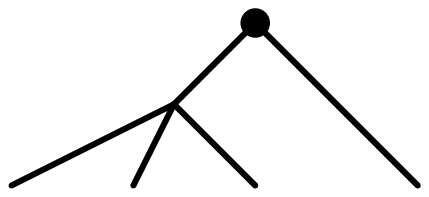}
             ~+~ \mbox{3 similar}
          \right]
       ~~~+~ \dots
\nonumber\\[0.04in]
   & & + \frac {12i\lambda }{(q^{2})^{2}}
             ~\rlap{\raisebox{0.20in}{\hspace*{0.65in}$\frac {1}{4!}\phi
^{4}$}}
             \epscenterbox{1.0in}{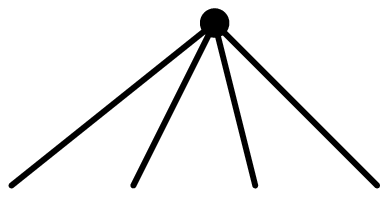}
       ~~~+~ \dots .
\end{eqnarray}
The dots indicate terms with other operators, for example, the
ones in Eq.\ (\ref{eq:tree2}), and again thick lines are those
that are forced to carry large momentum. Evidently the pattern is
general, but we must remember to add in all operators, not just
those with two $\phi $ fields.

\subsubsection{Loops}

The one-loop graphs for Eq.\ (\ref{eq:OPE}) are obtained from:
\begin{eqnarray}
   \epscenterbox{0.8in}{dis-sm.eps}
   &=& \epscenterbox{1.0in}{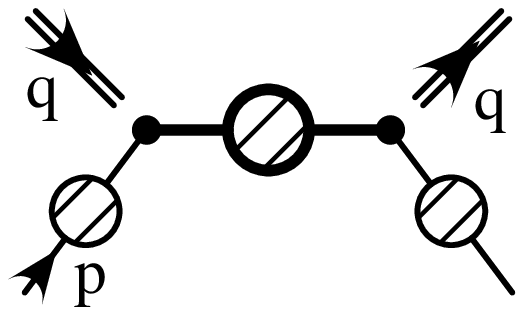}
       ~+~ \epscenterbox{1.0in}{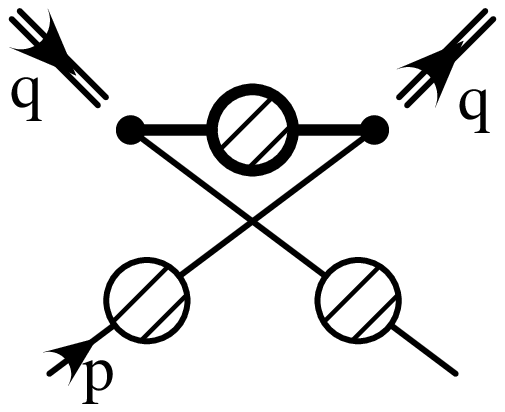}
\nonumber\\
   & & \hspace*{-1in}
       +~ \epscenterbox{0.9in}{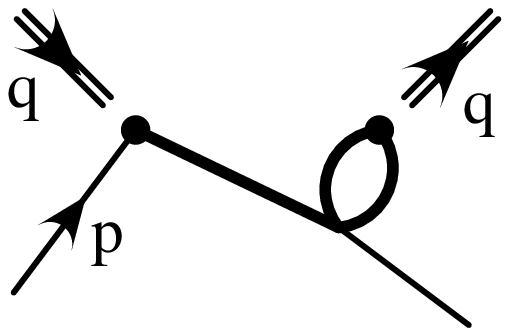} ~+~ \mbox{3 similar}
       ~~+~ \epscenterbox{0.9in}{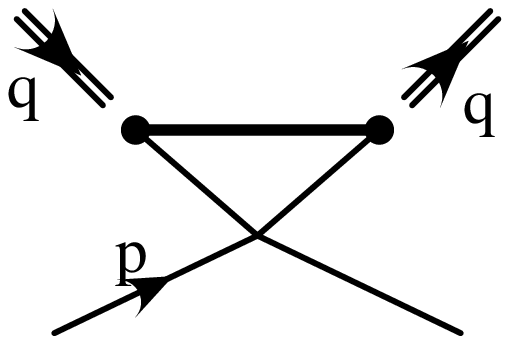} ~+~ O(\lambda ^{2}) .
\label{eq:loop2}
\end{eqnarray}
To economize on notation, I have exhibited the
graphs with propagator corrections by replacing free by complete
propagators everywhere in the tree graphs.  Let us examine only
the leading power ($1/q^{2}$), so that we only need the terms with
the $\phi ^{2}$ operator.  So we need to pick out the order $\lambda $ terms in
the OPE:
\begin{equation}
   C(q^{2}) ~~~
   \left[ ~
       \epscenterbox{0.8in}{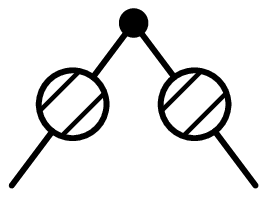} ~+~
       \epscenterbox{0.7in}{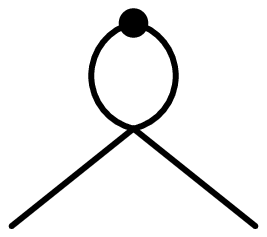} ~+~ O(\lambda ^{2})
   ~ \right] ,
\label{eq:OPEloop2}
\end{equation}
where we write the coefficient as the lowest order term, from
Eq.\ (\ref{eq:tree2}), times a power series in $\lambda $:
\begin{equation}
   C(q^{2}) = \frac {2i}{q^{2}} \left[ 1 + \lambda  C^{(1)}(q^{2}) + O(\lambda
^{2}) \right] .
\end{equation}

There are some obvious correspondences.  For example, graphs
in Eq.\ (\ref{eq:loop2}) with an external propagator correction
correspond to terms in Eq.\ (\ref{eq:OPEloop2}) with a
lowest-order coefficient times a propagator correction to the
lowest-order matrix element.

However, there are some clear differences.  For example, loops
including the vertex for the ``current'' $j$ can have
ultra-violet divergences, as in the first graph on the second
line of Eq.\ (\ref{eq:loop2}), while loops including the
operators in the OPE can have UV divergences, as in the second
graph in Eq.\ (\ref{eq:OPEloop2}).

Moreover, in the last graph in Eq.\ (\ref{eq:loop2}), we see a
non-trivial case where the loop momentum can range from being
soft to hard.  We cannot uniquely assign the lines to hard and
soft subgraphs. This is the source of the characteristic
complications of the derivation of the OPE.

But first we must understand how to define the operators; they
are ultra-violet divergent and need renormalization.

\subsection{Composite operators and their renormalization}

In the above equations, we have seen a number of examples of
composite operators, i.e., operators which are the product of
fields and their derivatives at the same space-time point.  There
are a number of situations where we need to treat matrix elements
and Green functions involving composite operators:
\begin{itemize}

\item  To represent the coupling of QCD to electroweak fields.

\item  For currents and charges, as in Noether's theorem.
    The operators for coupling QCD to electroweak gauge bosons
    are Noether currents.

\item  In the OPE, as auxiliary constructs.

\end{itemize}

These operators, in general, can have UV divergences beyond those
that are cancelled by counterterms in the QCD Lagrangian, except
in the case of properly defined Noether currents for symmetries.
The composite operators themselves need to be renormalized, by
following the same principles as for the renormalization of the
interactions.

\begin{figure}
    \[
       \begin{array}{*{7}c}
            & \epscenterbox{0.7in}{sf-2-op1.eps}
          &+& \epscenterbox{0.7in}{sf-2-op2.eps}
          &+& \epscenterbox{0.7in}{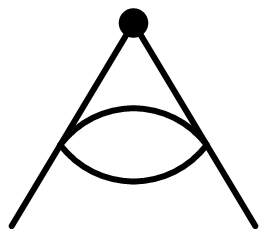}
          &+~~\dots
        \\
           & \mbox{(a)} && \mbox{(b)} && \mbox{(c)}
        \\[0.15in]
           +& \epscenterbox{0.7in}{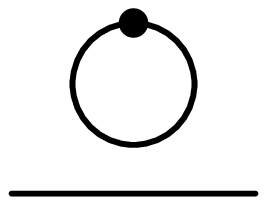}
          &+~~ \dots
        \\[0.35in]
           & \mbox{(d)}
        \end{array}
    \]
   \caption{Graphs for the $\phi \,\phi $ Green function of $\frac {1}{2}\phi
^{2}$. }
   \label{fig:CompOp}
\end{figure}

\begin{figure}
    \[
       \begin{array}{*5c}
            \epscenterbox{0.6in}{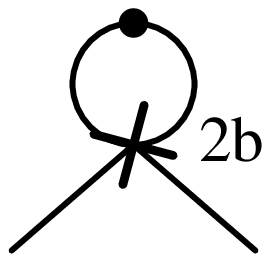}
          &~~& \epscenterbox{0.45in}{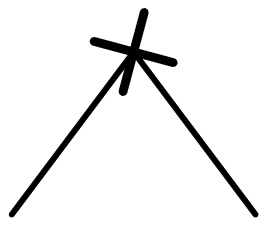}
          &~~& \epscenterbox{0.4in}{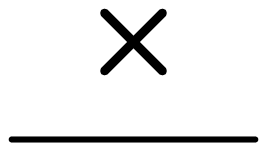}
        \\
           \mbox{(a)} && \mbox{(b)} && \mbox{(c)}
        \end{array}
    \]
   \caption{Counterterm graphs for Fig.\ \protect\ref{fig:CompOp}. }
   \label{fig:CompOpRen}
\end{figure}

Consider the graphs for
$\langle 0|T \half\phi ^{2} \, \tilde\phi (p_{1}) \, \tilde\phi (p_{2})
|0\rangle $, shown in Fig.\
\ref{fig:CompOp}.  Some of the divergences we have already seen.
For example, the one-loop subdivergence in (c) is cancelled by a
interaction counterterm that we have already calculated; this
gives the graph in Fig.\ \ref{fig:CompOpRen}(a). But some of the
other divergences are new.  These all include the vertex for the
$\frac {1}{2}\phi ^{2}$ operator, and they may be cancelled by extra
counterterms.
We now examine the interpretation of the new counterterms.

The counterterms are in fact of the form of vertices for
composite operators.  For example the counterterms for the
overall divergences in Fig.\ \ref{fig:CompOp}(b) and (c) are a
coefficient times the vertex for $\frac {1}{2}\phi ^{2}$, as shown in Fig.\
\ref{fig:CompOpRen}(b).  The disconnected graph Fig.\
\ref{fig:CompOp}(d) needs a counterterm proportional to the unit
operator, as in Fig.\ \ref{fig:CompOpRen}(c).

It should be evident that this is a general result: We can
generate renormalized Green functions of a composite operator by
including counterterms of the form of coefficients times
(vertices for) composite operators.  It follows that we may
define a renormalized $\frac {1}{2}\phi ^{2}$ operator as
\begin{equation}
   R \left[ \half\phi ^{2}\right]
   = \half\phi ^{2} [1 + \delta _{1} ] + 1 \, \delta _{2} \, m^{2} ,
\label{eq:RenPhi2}
\end{equation}
where the `1' stands for the unit operator, and $\delta _{1}(\lambda ,\epsilon
)$ and
$\delta _{2}(\lambda ,\epsilon )$ are the counterterm coefficients.  The usual
power
counting arguments tell us that in a renormalizable theory we
need as counterterms all operators of the same and lower
dimension as the operator we start with, modulo the restrictions
imposed by symmetries of the theory.  (In this case the relevant
symmetries are Lorentz invariance and $\phi \to -\phi $.)

On the right-hand side of the OPE Eq.\ (\ref{eq:OPE}) we must use
renormalized operators, so that we have finite quantities to work
with.  In effect, we, the users of the OPE, get to choose our
definitions of the operators, and then we must prove that the
coefficients can be chosen to make the OPE true.

\subsection{Construction of OPE}

In the first three graphs on the right-hand side of Eq.\
(\ref{eq:loop2}), we have some external propagators times a
subgraph through which large momentum flows.  So we may set
$p=m=0$ in the subgraph with the large momentum.  This gives a
$q$ dependent factor times the vertex for $\half\phi ^{2}$ times the
external propagators, i.e., we obtain purely a contribution to
the $O(\lambda )$ part of the coefficient.

The only remaining graph is in the last line of Eq.\
(\ref{eq:loop2}).  The loop has the value
\begin{equation}
   -\lambda  \int  \frac {d^{4}k}{(2\pi )^{4}} \frac {1}{( \BDpos k^{2}-m^{2}+i\epsilon
)^{2} \, \left[\BDpos(q+k)^{2}-m^{2}+i\epsilon \right]} .
\label{eq:LastLoop}
\end{equation}
First observe that in a region where $k$ is fixed as $q$ gets
large, we have a contribution
\begin{equation}
   \frac {2i}{q^{2}}  \frac {i\lambda }{2} \int _{\rm small} \frac
{d^{4}k}{(2\pi )^{4}} \frac {1}{( \BDpos k^{2}-m^{2}+i\epsilon )^{2}} .
\end{equation}
We have organized the normalization factors so that this is
manifestly the lowest-order coefficient times the one-loop Green
function of the $\half\phi ^{2}$ operator, the second graph in Eq.\
(\ref{eq:OPEloop2}).  This is a completely expected contribution.

The one-loop part of operator Green function also has a
contribution from large $k$, where the above argument does not
work.  The original graph has a contribution when $k$ is of order
$q$ or bigger.  To avoid double counting we take that graph and
subtract the one-loop Green function of $\half\phi ^{2}$ times the
lowest-order Wilson coefficient.  It is now legitimate to neglect
$p$ and $m$, so that we have
\begin{equation}
   -\lambda  \int  \frac {d^{4}k}{(2\pi )^{4}} \left\{
            \frac { \BDpos 1}{(k^{2})^{2} \, (q+k)^{2}}
            ~-~ \left[ \frac {1}{(k^{2})^{2}} - \mbox{UV counterterm} \right]
              \frac {1}{ \BDpos q^{2}}
      \right\} .
\end{equation}
This represents the original graph minus a corresponding
term that we already know is on the right of the OPE.  The effect
is that we have subtracted the low momentum behavior, and the
infra-red divergence that would otherwise be present when $m=p=0$
has cancelled.  The result it now an $O(\lambda )$ contribution to the
coefficient.

The generalization of the above results is that the $q\to \infty $ of the
left hand-side of Eq.\ (\ref{eq:OPE}) is a sum of terms with a
representation in Eq.\ (\ref{eq:OPEGraph}).  The $1/q^{2}$
coefficient of $\frac {1}{2}\phi ^{2}$ is a subtracted sum of graphs for
$\langle 0|T j(q) \, j \, \tilde\phi (p_{1}) \, \tilde\phi (p_{2}) |0\rangle $,
1PI in the $\tilde\phi $ legs, with $m$ and the $p$s set to zero, and
with subtractions made to cancel the resulting IR divergences.

\subsection{Proof}

It is not too hard to sketch a proof.  For each graph $\Gamma $ for the
left-hand-side of the OPE, we define a remainder
\begin{equation}
   r(\Gamma ) = \Gamma  - \mbox{UV counterterms}
          - \mbox{Subtractions for leading $q\to \infty $ behavior}.
\end{equation}
The subtractions for the large $q$ behavior can be constructed in
much the same fashion as renormalization counterterms.  Much of
the technology is very similar, in fact.  If we just subtract the
leading power ($1/q^{2}$ times logarithms of $q$), then the
remainder is of order $1/q^{3}$ times logarithms.  Subtracting more
terms gives a correspondingly smaller remainder.  For the 1-loop
graph we just considered, we write
\begin{eqnarray}
  r\left(\mbox{1 loop graph}\right)
  &=& \mbox{1 loop graph}
    + C(\mbox{tree})  \, R\left[ \mbox{1 loop ME} \right]
\nonumber\\
  & & + C(\mbox{1 loop graph}) ,
\end{eqnarray}
where now $C(\Gamma )$ means that we take the negative of the subgraph
$\Gamma $ with $m=p=0$, {\em after subtraction of IR divergences}.

There are of course the usual difficult issues of showing that
the subtractions actually cancel the divergences, and that the
standard power-counting criteria are correct; in particular that
the remainder is actually suppressed by the claimed power of $q$
compared with the original graphs.

Then we sum over all graphs and show that the subtraction terms
have the form of coefficients times Green functions of operators.

\subsection{Use of renormalization group}

Consider the leading term in the OPE Eq.\ (\ref{eq:OPE}) for
connected graphs.  Let us write it as
\begin{equation}
   C(q^{2}, \lambda (\mu ), \mu ) \,
   \,   G_{\frac {1}{2}\phi ^{2}}(p, \lambda (\mu ), \mu , m) .
\end{equation}
When $q\to \infty $ with the other variables fixed, we get logarithms of
$q/\mu $ in higher order contributions to the coefficient $C$.
These ruin the accuracy of fixed-order calculations.

But we may use the renormalization group to set $\mu =|q|$.
We can then use a low order calculation of
$C(q^{2}, \lambda (|q|), |q|)$, provided only that $\lambda (|q|)/16\pi ^{2}$
is small.
A knowledge of the anomalous dimension of the fields and of $\frac {1}{2}\phi
^{2}$
enables us to express $G_{\frac {1}{2}\phi ^{2}}$ at $\mu =|q|$ in terms of its
value at
fixed $\mu $.

The important facts are that we have separated the original Green
function, which depends on two widely different scales $q$ and
$m$ (or $p$), into two factors, each of which depends on only one
of the scales.  We can now change the renormalization mass $\mu $
independently in each factor, and a change of $\mu $ suffices to
remove the logarithms. It is essential for this that the Wilson
coefficient $C$ have no IR divergences, after subtraction.

If we left $\mu $ fixed we would have a series of the form
\begin{eqnarray}
   C &=& \frac {{\rm constant}}{q^{2}}
         \Bigg[
            1 + \lambda  \left( \# \ln \frac {q^{2}}{\mu ^{2}} + \# \right)
\nonumber\\
     & & ~~~~~~~~~~
           \lambda ^{2} \left( \# \ln^{2} \frac {q^{2}}{\mu ^{2}} + \# \ln
\frac {q^{2}}{\mu ^{2}} + \# \right)
           + \dots \Bigg] ,
\end{eqnarray}
where, the `$\#$'s stand for numerical constants.

\subsection{Renormalization group equation}

By expressing the renormalized operators in terms of bare
operators, we can find RG equations
\begin{equation}
   \mu \frac {d\phi }{d\mu } = \mu \frac {dZ^{-1/2}}{d\mu } \, \phi _{0} =
-\half \gamma (\lambda (\mu )) \, \phi  ,
\end{equation}
\begin{equation}
   \mu \frac {dR[\phi ^{2}]}{d\mu } = \mu \frac {dZ_{\phi ^{2}}}{d\mu } \, \phi
_{0}^{2} = \gamma _{\phi ^{2}}(\lambda (\mu )) \,\, R[\phi ^{2}],
\end{equation}
with
\begin{equation}
   \gamma _{\phi ^{2}}(\lambda (\mu )) = \beta (\lambda ,\epsilon ) \frac
{\partial  \ln Z_{\phi ^{2}}(\lambda ,\epsilon )}{\partial \lambda } .
\end{equation}
Here, $Z_{\phi ^{2}}$ is the renormalization factor of the $\phi ^{2}$
operator:
$Z_{\phi ^{2}} = (1+\delta _{1})Z^{-1}$ in the notation of Eq.\
(\ref{eq:RenPhi2}).  We
have ignored, for simplicity, the term with the unit operator; it
is not needed if we treat connected Green functions only.

Similar equations can be derived for the other composite
operators.

{}From these equations follows a renormalization group equation for
the Wilson coefficient:
\begin{equation}
   \mu \frac {d}{d\mu } C(q^{2}, \lambda (\mu ), \mu )
   = \left[ 2 \gamma _{j}(\lambda ) - \gamma _{\phi ^{2}}(\lambda ) \right] \,
C ,
\end{equation}
where $\gamma _{j}$ is the anomalous dimension of the ``current'' $j$,
which in our simplified example happens also to be the operator
$\phi ^{2}$.

The final form of the leading term for the large $q$ behavior of
the left-hand side of Eq.\ (\ref{eq:OPE}) is
\begin{equation}
   e ^{\int _{\mu }^{|q|} \frac {d\mu '}{\mu '} \left[ - 2 \gamma _{j}(\lambda
(\mu ')) + \gamma _{\phi ^{2}}(\lambda (\mu ')) \right]}
   \,\, C(q^{2},\lambda (|q|),|q|)
   \,\, G_{\frac {1}{2}\phi ^{2}}(p, \lambda (\mu ), \mu , m) .
\end{equation}

Consequences of this result and its generalizations include
\begin{itemize}

\item In a theory with a non-zero UV fixed point, we obtain a
    power law not equal to the na\"\i ve one, but we cannot do
    perturbative calculations.

\item In an asymptotically free theory, like QCD, we obtain the
    power law given by dimensional analysis, but with logarithmic
    corrections.  The errors are under control, since we can
    expand the anomalous dimensions and the Wilson coefficients
    in powers of a small parameter $\alpha _{s}$.  The operator matrix
    elements cannot be calculated perturbatively; at the present
    state of the art, they must be measured experimentally.
    However, the matrix elements are universal: the same ones
    can be used in treating many different processes.

\item If $j$ is a conserved current (e.g., an electromagnetic
    current), then its anomalous dimension is exactly zero,
    $\gamma _{j}=0$.

\end{itemize}



\section{Renormalization etc in the standard model}
\label{sec:other.theories}

In this section, I will summarize the issues that arise in
generalizing the results of the previous sections to the standard
model, and indeed to a general quantum field theory.  Among these
issues are
\begin{itemize}

\item  There are a lot of extra couplings.  Is the theory
    nevertheless renormalizable?  What are the RG equations?

\item  There are many symmetries, some of them broken.  How do we
    treat renormalization, the RG and the OPE in the presence of:
    global symmetries, local symmetries, spontaneous symmetry
    breaking?  What are the special problems that arise with
    chiral symmetries?

\item  In particular there is the possibility of ``anomalous
    breaking'' of symmetries.

\item  A particular issue of importance in QCD is to determine
    which operators we need to use in the OPE.  It needs a highly
    non-trivial proof to show that we can omit all but
    gauge-invariant operators.

\end{itemize}
I will only have space to present a summary of the issues.

The Lagrangian for the standard model, with its
$SU(3)\times SU(2)\times U(1)$ symmetry, can be written in a very compact
form:
\begin{equation}
   {\cal L} = -\quarter {G_{\mu \nu }^{\alpha }}^{2}
            + \bar\psi  (i \st{D} - M ) \psi
            + D^{\mu }\phi ^{\dagger }D^{\mu }\phi
            - V(\phi ,\phi ^{\dagger }) + \bar\psi  \Gamma \psi \phi  + h.c.
\end{equation}
Here we use a standard notation, for the gauge, fermion and Higgs
fields.  All these fields have been put together into big
vectors, so that there are implicitly a lot of summed indices in
each term.

I will address in turn the different issues involved in
renormalizing this theory by examining a series of simplified
model theories.  The issues are generic to quantum field
theories, and have nothing to do with the specific form of the
standard model.

\subsection{Need for extra interactions}

Consider a Yukawa model of a Dirac field and a real scalar field:
\begin{equation}
   {\cal L}_{\rm v1} = \BDpos \half\partial \phi^{2} -\half m^{2}\phi^{2}
        + \bar\psi  (i \st{\partial } - M ) \psi
        + g\bar\psi \psi \phi .
\label{eq:Yuk1}
\end{equation}
We start off by finding the divergent one-loop graphs.

\begin{figure}
    \begin{center}
        \leavevmode
        \epscenterbox{0.8in}{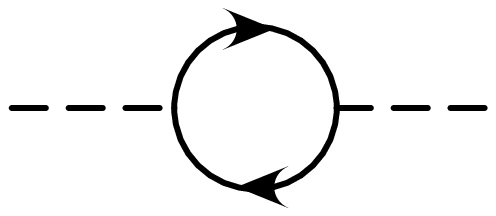}
        ~~~~ \epscenterbox{0.9in}{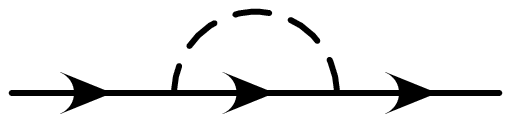}
        ~~~~ \epscenterbox{0.9in}{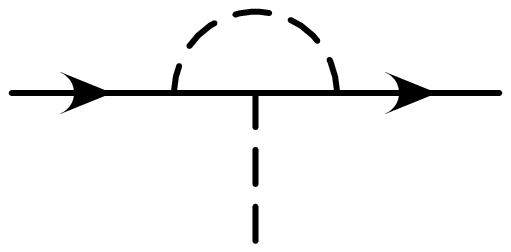}
    \end{center}
    \caption{Divergent one-loop graphs in Yukawa theory that have
             counterterms formed by renormalizing the original
             Lagrangian. }
\label{fig:YukDiv1}
\end{figure}

\begin{figure}
    \begin{center}
        \leavevmode
        \epscenterbox{0.6in}{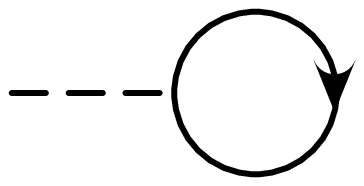}
        ~~~~~ \epscenterbox{0.8in}{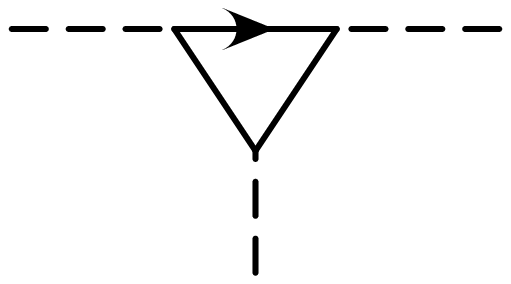}
        ~~~~ \epscenterbox{0.8in}{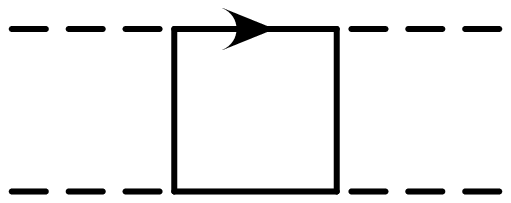}
    \end{center}
    \caption{Divergent one-loop graphs in Yukawa theory that do
             not have counterterms corresponding to terms in the
             original Lagrangian. }
\label{fig:YukDiv2}
\end{figure}

The scalar self energy, the fermion self energy and the
fermion-scalar vertex (Fig.\ \ref{fig:YukDiv1}) have degrees of
divergence 2, 1, 0, and thus their divergences can be cancelled
by counterterms proportional to terms in Eq.\ (\ref{eq:Yuk1}),
just as in the $\phi ^{4}$ theory.  But in addition there are other
divergent graphs, shown in Fig.\ \ref{fig:YukDiv2}.  These need
counterterms proportional to $\phi $, $\phi ^{3}$, and $\phi ^{4}$, terms not
in
Eq.\ (\ref{eq:Yuk1}).  Note that the first two graphs have
degrees of divergence above zero, but there are no
Lorentz-invariant operators of an appropriate dimension except
the ones listed.

We cancel all these divergences by renormalizing the original
Lagrangian and then adding some extra terms:
\begin{eqnarray}
   {\cal L}_{\rm v2} &=& \BDpos \half Z_{\phi } \partial \phi ^{2} - \half
m_{B}^{2}\phi ^{2}
        + Z_{\psi } \bar\psi  i \st{\partial } \psi  - M_{B} \bar\psi  \psi
        + g_{B}\bar\psi \psi \phi
\nonumber\\
        & & - \lambda _{B1}\phi  - \lambda _{B3}\phi ^{3} - \lambda _{B4}\phi
^{4} .
\label{eq:Yuk2}
\end{eqnarray}
Since we have added 3 new couplings, we must ask ``Does this
process ever stop? Do we have to keep adding even more couplings
as we go to higher order?''

The answer is that no extra couplings are needed.  We notice
first that at one loop order, graphs with more external lines
than those we listed are convergent.  Then we observe that not
only the original couplings but also the extra ones have a mass
dimension that is zero or larger.  Let us apply dimensional
analysis.  The degree of UV divergence of a graph equals the
dimension of its integral, so that
\begin{equation}
   {\mbox{dim.\ of graph} \atop D(\Gamma )}
   ~=~
   {\mbox{dim.\ of couplings} \atop \delta (\Gamma )}
   ~+~
   {\mbox{deg.\ of divergence} \atop \Delta (\Gamma )}
\end{equation}
The degree of divergence is therefore $D(\Gamma )-\delta (\Gamma )$.  The
counterterm is a polynomial in momenta of this degree. Now, the
couplings in the counterterms are the coefficients of the terms
in the polynomial. Since the whole counterterm polynomial is to
be added to the graph, its dimension is the same as that of the
graph, i.e., $D(\Gamma )$. Hence the dimensions of the couplings of the
counterterms range from $D(\Gamma )$, for a term that is independent of
momentum, down to $D(\Gamma )-\Delta (\Gamma )=\delta (\Gamma )$, for a term
with the maximum
power of momentum.  Hence the dimensions of the counterterm
couplings are all at least as large as $\delta (\Gamma )$, which is the total
of the dimensions of the couplings in the original graph.

Hence if we start with couplings of non-negative dimension, all
the counterterm couplings have non-negative dimension. There are
a finite number of such terms, since the fields have positive
dimension, and thus the process of generating new counterterm
vertices terminates.  Indeed, in the second version of the Yukawa
Lagrangian, Eq.\ (\ref{eq:Yuk2}), we have all possible terms with
couplings of non-negative dimension, given that they must be
Lorentz invariant and parity-invariant.  The last requirement
prohibits terms with $\gamma _{5}$s in them, and is obeyed because the
original Lagrangian is parity-invariant.

One important result follows, that if we write down a candidate
Lagrangian for a theory of physics, then we may be forced to
extend it by the addition of extra terms in order to get a
renormalizable theory.

\subsection{General results on renormalizability}

We have the following cases
\begin{itemize}

\item  All couplings in the initial candidate Lagrangian have
    dimension $\geq 0$.  Then either
    \begin{quote}
        No new couplings are needed, as in the $\phi ^{4}$ theory.
    \end{quote}
    or
    \begin{quote}
        We need new couplings, all of dimension $\geq 0$.  But the
        number of such terms is finite.
    \end{quote}
    Such theories are called ``renormalizable'' or
    ``renormalizable with the addition of extra couplings''.

\item At least one coupling has dimension $<0$.  Then for each
    Green function, we can get a divergence of arbitrarily high
    degree by going to a high enough order of perturbation
    theory.  We need an infinite collection of counterterms, and
    such theories are termed ``non-renormalizable''.  The
    classic physical case is the 4-fermion theory of weak
    interactions.

\end{itemize}
A special case of a renormalizable theory is a
super-renormalizable theory, where only a finite number of 1PI
{\em graphs} need counterterms for overall divergences.  This
happens when all of the interactions have strictly positive
dimension, as in $\phi ^{4}$ theory in less than four space-time
dimensions, but it does not happen in any interacting four
dimensional theory.\footnote{
    An apparent counterexample is $\phi ^{3}$ theory.  However,
    its energy density is not bounded below, and the theory is
    unstable against decay of the vacuum; it is therefore
    unphysical.  However, the theory is a useful source of
    exercises on perturbation theory, where the difficulties are
    not manifest.
}

It is also possible to get zeros in the coefficients of
divergences, so that fewer counterterms are needed than indicate
by the power-counting diagnosis.  This is a property of specific
theories, particularly supersymmetric theories.

It is also possible that the situation may change beyond
perturbation theory.  Even so, very powerful results are needed
to convert an theory that is apparently non-renormalizable theory
by the power-counting criterion to a genuinely renormalizable
theory. Our experience with non-renormalizable theories of
physics is restricted to weak interactions and to gravity.  In
the first case the non-renormalizable theory is an approximation
to the true theory, but only in a certain domain.  In the second
case, we still have no complete, accepted and useful theory of
quantum gravity.

\subsection{Renormalization group and OPE with many couplings}

The principles of the renormalization group that I explained in
Sect.\ \ref{sec:RG} can be applied to any field theory.  The main
difference is that we obtain a $\beta $ function for each coupling,
and that it is generally a function of all the couplings.  Thus
the equation for the running of the coupling becomes an array of
equations.  For example in the Yukawa theory Eq.\
(\ref{eq:Yuk2}), we have
\begin{equation}
   \mu \frac {d}{d\mu } \left(
        \begin{array}{c}
             \lambda _{4} \\
             g
        \end{array}
       \right)
   =
   \mu \frac {d}{d\mu } \left(
        \begin{array}{c}
             \beta _{4}(\lambda _{4},g) \\
             \beta _{g}(\lambda _{4},g)
        \end{array}
       \right).
\end{equation}
(The dimensional coupling $\lambda _{3}$ does not enter into the RGEs for
the dimensionless couplings.)
Similar equations can be written for the dimensional couplings
and the masses, and for the anomalous dimensions.

Anomalous dimensions for the fields become functions of the
several couplings, as do anomalous dimensions for composite
operators.

Furthermore, when one obtains the OPE in a general theory one
must consider all operators of the appropriate dimension, and the
renormalization-group equation will mix all the operators of a
given dimension.  But once that is realized, all the same
principles apply as in the $\phi ^{4}$ theory.

\subsection{The physical significance of non-renormalizable
    theories}

The above discussion appears to indicate that non-renormalizable
theories are inadmissible as theories of physics, on the grounds
that one must introduce an infinite collection of counterterms to
renormalize them.  The inherent arbitrariness in choosing the
finite parts of counterterms then implies that an infinite set of
parameters is needed to specify the theory.

However, this assumes that a field theory must be used with the
UV cut-off removed.  However, all that is necessary is that the
momentum scale of the cut-off is sufficiently much bigger than
the scales currently accessible to experiments. More generally,
one may suppose that there is some true fundamental theory of
physics, and all that one can directly observe are low energy
approximations to this true theory.

\begin{figure}
    \begin{center}
       \leavevmode
       $\epscenterbox{0.6in}{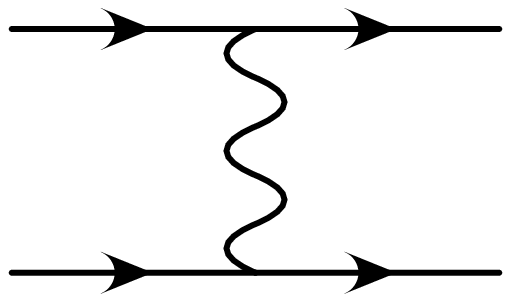}
       ~\to ~ \epscenterbox{0.6in}{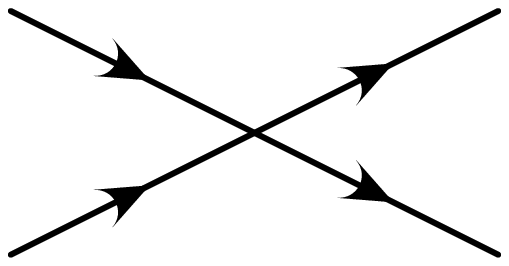}
       $
    \end{center}
    \caption{A 4-fermion interaction is a low energy approximation
            to $W$-exchange.  }
\label{fig:WI}
\end{figure}

The classic example are the semi-leptonic weak interactions, as
illustrated in Fig.\ \ref{fig:WI}.  At low energies, the $W$'s
propagator can be approximated by a constant: $1/(\BDpos q^{2}-m_{W}^{2}) \to
-1/m_{W}^{2}$.  Thus one obtains an effective 4-fermion coupling
$G_{F} \propto  g^{2}/m_{W}^{2}$.  The negative dimension of $G_{F}$ is exactly
the
signature of a non-renormalizable theory.
Since an exchange of a light particle,
like the photon, will give a much larger propagator, weak
interaction amplitudes are suppressed by a power $q^{2}/m_{W}^{2}$.

{}From this point of view, a characteristic of non-renormalizable
interactions is that in the region where they are a good
approximation to reality they are weak.  Perturbation theory in
the non-renormalizable coupling can be considered as an expansion
in powers of the typical momentum scale of a quantity being
calculated divided by a large scale like $M_{W}$.  The large scale
is effectively a cut off on the non-renormalizable theory.

\begin{figure}
    \begin{center}
       \leavevmode
       \epscenterbox{0.8in}{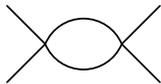}
    \end{center}
    \caption{Quadratically divergent graph in 4-fermion theory.}
\label{fig:WIDiv}
\end{figure}

The divergences in higher order graphs for a non-renormalizable
theory, as in Fig.\ \ref{fig:WIDiv}, come from regions of large
momentum.  That is precisely where the 4-fermion vertices are bad
approximations to the true theory.  We can choose to add
counterterms to renormalize the graph, and the finite parts of
the counterterms will be determined by the true theory.  After
the counterterms are added, perturbation theory for the
non-renormalizable theory is an expansion in powers of $q^{2}/m_{W}^{2}$.
The arbitrariness in the finite parts affects higher order terms
in this expansion, so we are not directly bothered by the
infinite number of the counterterms.

If a non-renormalizable theory is a theory of physics, then it
indicates a momentum scale at which it breaks down, which is
where its interactions stop being small corrections to
renormalizable interactions.  For the 4-fermion theory, this is
of the order of 100~GeV, which is exactly where the
Weinberg-Salam theory must be used.

The other established non-renormalizable theory of physics is
general relativity, the theory of gravity.  The scale at which it
must breakdown is of the order of Planck mass.  This is far above
all experimentally accessible scales, and the quantum effects of
general relativity are negligible in normal situations, for
example in atomic and nuclear physics.  Gravity is intrinsically
many orders of magnitude weaker than even the weak interactions
at energy scales of a GeV.

Of course, gravity dominates the physics of large scale systems,
like the earth, the solar system and the universe.  But in these
cases we are dealing only with the classical-field limit of
gravity; quantum gravity is irrelevant.  For the long range
forces in large systems, strong and weak interactions are
unimportant because they have a finite range, and electromagnetic
interactions are unimportant because non-zero electric charges
are screened in large systems.

A renormalizable theory, in contrast to a non-renormalizable
theory, does not contain indications of its own breakdown; it is
a self-consistent theory when the UV cut-off is removed.  (To be
fair there are indications that non-asymptotically free theories
may not exist: For example, the bare coupling in QED becomes
strong if the UV cut-off is at about the Planck scale; this, at
the least, implies that perturbation theory is inadequate to
treat the continuum limit, unlike the case of an asymptotically
free theory.)

\subsection{Symmetries}

A theory like the standard model has many symmetries, and it is
necessary to prove, as far as possible, that the only
counterterms that are necessary to renormalize the theory
preserve the symmetry.

\begin{figure}
    \begin{center}
       \leavevmode
       \epscenterbox{0.8in}{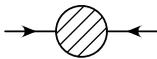}
    \end{center}
    \caption{These graphs are zero in the theory specified by Eq.\
             (\protect\ref{eq:ComplexPhi4}).  }
\label{fig:2in}
\end{figure}

A simple case is a global symmetry, such as the symmetry $\phi \to \phi
e^{i\omega }$
of the following Lagrangian of a complex scalar field:
\begin{equation}
   \partial \phi ^{\dagger } \partial \phi  - m^{2}\phi ^{\dagger }\phi  -
\lambda (\phi ^{\dagger }\phi )^{2}.
\label{eq:ComplexPhi4}
\end{equation}
By Noether's theorem this symmetry implies the existence of a
conserved abelian charge, together with an associated conserved
current.  It is not hard to show that all the counterterms needed
are symmetric.  For example, no mass counterterm proportional to
$\phi ^{2}$ is needed (as opposed to $\phi ^{\dagger }\phi $).  The simplest
proof is to
observe that the free propagator for a complex field is directed:
it has an arrow on it.  The interaction has two lines entering
and two lines leaving it, and so Green functions are zero if they
have unequal numbers of lines entering and leaving them, like
Fig.\ \ref{fig:2in}.

It is also possible to show that the same result on the
counterterms is true if there is spontaneous breaking of the
symmetry, as would occur if $m^{2}$ were negative. The simplest
proofs of these theorems are made by observing that the
symmetries are still valid when a UV regulator is applied to the
theory.   Corresponding result hold to restrict the operators
that appear in the OPE.

But chiral symmetries are not necessarily preserved.  There is a
theorem that normal UV regulators break them.  For example, one
cannot define a Dirac $\gamma _{5}$ matrix for a dimensionally regulated
theory without either losing some properties of the 4-dimensional
$\gamma _{5}$ that are needed to prove chiral invariance or losing
consistency.  So chiral symmetries can only be preserved, if at
all, in the limit that the regulator is removed.  In general, one
expects ``anomalies''. There are known criteria for the
cancellation of such anomalies, and one constraint of the fermion
content of the standard model is that anomalies cancel; the weak
and electromagnetic part of the gauge group is chiral, since the
weak bosons couple differently to left- and right-handed
fermions.

\subsection{Gauge theories}

Consider QED, whose Lagrangian, with a gauge -fixing term, is
\begin{equation}
   {\cal L} = -\quarter F_{\mu \nu }^{2}
            + \bar\psi  (i \st{D} - M ) \psi
            - \frac {1}{2\xi }\partial \cdot A^{2} ,
\label{eq:QED}
\end{equation}
where $F_{\mu \nu }=\partial _{\mu }A_{\nu }-\partial _{\nu }A_{\mu }$ and
$D_{\mu }\psi  = \partial _{\mu }\psi +ieA_{\mu }\psi $.
Apart from the gauge-fixing term, which we have chosen to be
$-\partial \cdot A^{2}/2\xi $, the Lagrangian is invariant under the local
transformation
\begin{eqnarray}
   A_{\mu }(x) &\to & A_{\mu }(x) + \partial _{\mu }\omega (x) ,
\nonumber\\
   \psi (x) &\to & e^{-ie\omega (x)} \psi (x).
\end{eqnarray}

Now, the power counting criterion allows counterterms proportional
to ${A_{\mu }}^{2}$ and to $({A_{\mu }}^{2})^{2}$, as well as to those in the
original
Lagrangian; i.e., it allows a photon mass term and a photon
self-interaction. These terms are not gauge invariant, and in
fact they are not generated as counterterms. But since the
Lagrangian after gauge fixing is not gauge-invariant, a simple
appeal to gauge invariance is not sufficient to provide aproof.

Even so, it can be shown that QED can be renormalized by the
following transformation of Eq.\ (\ref{eq:QED}):
\begin{equation}
   {\cal L} = - Z_{3} \quarter F_{\mu \nu }^{2}
            + Z_{2} \bar\psi  (i \st{D} - M_{0} ) \psi
            - \frac {1}{2\xi }\partial \cdot A^{2} .
\label{eq:QED.Ren}
\end{equation}
No renormalization of the gauge fixing term is needed, and the
covariant derivative is unchanged, if expressed in terms of the
renormalized fields and couplings.  It can be expressed in terms
of the bare fields: $D_{\mu }\psi  = \partial _{\mu }\psi +ieA_{\mu }\psi  =
\partial _{\mu }\psi +ie_{0}A^{(0)}_{\mu }\psi $, where the
bare coupling is $e_{0}=Z_{3}^{-1/2}e$.

After renormalization, the Lagrangian with the gauge-fixing term
omitted is still gauge-invariant.

A number of other results are needed in order to demonstrate that
the theory is a valid theory.  The basic problem is that there
are unphysical degrees of freedom and it must be proved that they
decouple from all of the genuine physics.  The proofs rely on
gauge invariance, so that the proof that renormalization
preserves gauge invariance is necessary if we are to prove that
the unphysical degrees of freedom continue to decouple after
renormalizing the theory.

First we define the concept of a {\em physical state}.  For
scattering states this is one where the photon polarizations obey
$k\cdot \epsilon =0$.  For a single photon, this still allows 3 polarizations
per momentum state, so in addition we identify states that differ
by a gauge transformation.  That is, we consider equivalent
states which differ by changing the polarization of a photon
state by $\epsilon _{\mu } \to  \epsilon _{\mu }+ak_{\mu }$.  (The number $a$
is arbitrary and may
differ from photon to photon.)

It is necessary to show that no contribution to physical matrix
elements is made by photons with scalar polarization ($\epsilon _{\mu }\propto
k_{\mu }$).
It is also necessary to prove unitarity.  That is, if ${\cal A}$
and ${\cal B}$ are any two physical operators and
$|{\rm phys 1}\rangle $ and $|{\rm phys 2}\rangle $ are any physical states,
then
\begin{equation}
   \langle {\rm phys \, 1}| {\cal A} {\cal B} |{\rm phys \, 2}\rangle
  = \sum _{n{\rm\ physical}} \langle {\rm phys \, 1}| {\cal A} |n\rangle
                      \langle n| {\cal B} |{\rm phys \, 2}\rangle  .
\end{equation}
Finally, it must be shown that physical matrix elements are
independent of the method of gauge-fixing, which implies that
they are independent of the arbitrary gauge-fixing parameter $\xi $.
A physical matrix element is a matrix element of a gauge
invariant operator between physical states.  One kind of physical
matrix element is an $S$-matrix element between physical states.

To show these results, we need the Ward identities.
Interestingly, these results continue to be true if a photon mass
is added to the Lagrangian, even though that breaks gauge
invariance.

It can also be shown that the OPE on physical states needs only
gauge-invariant operators:
\begin{equation}
   \langle {\rm phys \, 1}| \tilde j(q) j(0) |{\rm phys \, 2}\rangle
  = \sum _{i} C_{i}(q) \langle {\rm phys \, 1}| {\cal O}_{i} |{\rm phys \,
2}\rangle  ,
\end{equation}
where the sum is restricted to gauge-invariant operators
${\cal O}_{i}$.  It can be shown that other operators do appear in
the OPE, but that when we restrict to physical states they give
zero.

\subsection{Non-abelian gauge theories}

Similar results can be proved in non-abelian gauge theories,
including QCD by itself, as well as the standard model.  The
proofs get much harder, partly because of the presence of
Faddeev-Popov ghost fields in the gauge-fixed Lagrangian.
Although the Lagrangian is not gauge-invariant, it is invariant
under a kind of global supersymmetry, BRST symmetry.  Since BRST
transformations are non-linear, the methods that apply to
ordinary global symmetries do not work without modification.

One complication is that the gauge transformations are
renormalized, unlike the case of QED.  The covariant derivative
is written $D_{\mu }\psi  = \partial _{\mu }\psi +ig_{0}A^{(0)\alpha }_{\mu
}t_{\alpha }\psi $, where $t_{\alpha }$ are the
generating matrices for the gauge group acting on the matter
field.  But no longer do we have $g_{0}$ equal to $Z_{3}^{-1/2}g$.

\subsection{Summary}

\begin{itemize}

\item The key property that makes a theory renormalizable is the
    that is has couplings of non-negative dimension:
    ${\rm dim}({\rm couplings}) \geq  0$.

\item Ordinary global symmetries of the Lagrangian are preserved
    under renormalization, even when there is spontaneous
    breaking of the symmetry by the vacuum.

\item But this is not true for chiral symmetries unless anomaly
    cancellation conditions are satisfied.

\item Gauge-invariance is preserved under renormalization, but
    the proof is highly non-trivial.  Gauge-invariance is vital
    to the proof of unitarity and of other properties of physical
    states.

\end{itemize}



\section*{Acknowledgments}

My work was supported in part by the U.S. Department of Energy
under grant number DE-FG02-90ER-40577.


\appendix

\section{Useful integrals etc.}


\paragraph{The Gamma function} is defined by
\begin{equation}\Gamma (z)=\int _{0}^{\infty
}dt\;t^{z-1}\;e^{-t}.\end{equation}
It satisfies:
\begin{equation}\Gamma (z+1)=z\Gamma (z).\end{equation}
When $z=n>0$, an integer, $\Gamma (n)=(n-1)!$.  $\Gamma (z)$ has poles at all
negative and zero values of $z$.  The Taylor series expansion about
$z=0$ is
\begin{equation}\Gamma (z)=\frac {1}{z}\,\left[1-\gamma _{E}z+O(z^{2})\right
],\end{equation}
where $\gamma _{E}=0.5772...$ is Euler's constant.  Also
\begin{equation}\Gamma (1/2)=\pi ^{1/2}.\end{equation}

\paragraph{The Beta function} is defined by
\begin{equation}
    B(\alpha ,\beta ) = \frac {\Gamma (\alpha )\Gamma ( \beta )}{\Gamma (\alpha
+\beta )}
    = \int _{0}^{1}dx \, x^{\alpha -1}(1-x)^{\beta -1} .
\end{equation}
Hence:
\begin{eqnarray}
   \int _{0}^{\infty }dx \, x^{\alpha }\,(x+A)^{\beta }
   &=& A^{1+\alpha +\beta } B(\alpha +1,-\beta -\alpha -1)
\nonumber\\
   &=&A^{1+\alpha +\beta }\frac {\Gamma (\alpha +1)\Gamma (-\beta -\alpha
-1)}{\Gamma (-\beta )}.
\end{eqnarray}

\paragraph{Feynman parameters:}
\begin{equation}
   \prod _{j=1}^{N}\frac {1}{A_{j}^{\alpha _{j}}}
   = \prod _{j=1}^{ N}\left(\int _{0}^{1}dx_{j}\;x_{j}^{\alpha _{j}-1}\right)
   \frac {\Gamma \!\left(\sum _{ j}\alpha _{j}\right)}{\prod _{j}\Gamma (\alpha
_{j})}
   \;\frac {\delta \left(1-\sum _{ j=1}^{N}x_{j}\right)}{\left(\sum
_{j=1}^{N}A_{j}x_{j}\right)^{\sum \alpha _{ j}}} ,
\label{eq:Feyn.param}
\end{equation}

\paragraph{Spherically symmetric $n$-dimensional Euclidean
integrals:}
\begin{equation}
   \int d^{n}k\;f(|k|)=\frac {2\pi ^{n/2}}{\Gamma (n/2)}\;
   \int _{0}^{\infty }dk\;k^{n-1}\;f(k).
\label{eq:sph.sym}
\end{equation}

\paragraph {Minkowski space integrals:}
\emph{All with the metric where
  $a\cdot b = \BDpos a^0b^0 \BDminus {\bf a}\cdot{\bf b}$.
}
\begin{equation}
   \int d^{n}k \; \frac {1}{\left(\BDneg k^{2} \BDminus 2p\cdot k+C\right)^{\alpha }}
    = \frac { i\pi ^{n/2}\;\Gamma (\alpha -n/2)}{\Gamma (\alpha )}
      \,\left(C \BDplus p^{2}\right)^{n/2-\alpha },
\label{eq:MSI0}
\end{equation}
\begin{equation}
   \int d^{n}k\;\frac {k^{\mu }}{\left(\BDneg k^{2} \BDminus 2p\cdot k+C\right)^{\alpha }}=
   \frac {i\pi ^{n/2}\;\Gamma (\alpha -n/2)}{\Gamma (\alpha )}
   \,\left(C \BDplus p^{2}\right)^{n/2-\alpha }\,(-p^{\mu }),
\label{eq:MSI1}
\end{equation}
\begin{eqnarray}
   \int d^{n}k\;\frac {k^{\mu }k^{\nu }}
                 {\left(\BDneg k^{2} \BDminus 2p\cdot k+C\right)^{\alpha }}
    &=& \frac {i\pi ^{n/2}}{\Gamma (\alpha )}\,\left(C \BDplus p^{2}\right)^{n/2-\alpha
}
\nonumber
\\
&&
   \hspace*{-1.1in}
   \left[\Gamma (\alpha -n/2)p^{\mu }p^{\nu } 
         \BDminus \frac {1}{2}\Gamma (\alpha-1-n/2)g^{\mu \nu }( C \BDplus p^{2} )
   \right],
\label{eq:MSI2}
\end{eqnarray}
\begin{eqnarray}
   \int d^{n}k\;\frac {k^{2}}
                 {\left(\BDneg k^{2} \BDminus 2p\cdot k+C\right)^{\alpha }}
   &=&\frac { i\pi ^{n/2}}{\Gamma (\alpha )} \,
    \left(C \BDminus p^{2} \right)^{n/2-\alpha}
\nonumber\\
&&
   \hspace*{-1in}
   \left[ \Gamma(\alpha -n/2) p^{2}
           \BDminus \frac {n}{2}\Gamma (\alpha-1-n/2)( C \BDplus p^{2} )
   \right].
\label{eq:MSI2a}
\end{eqnarray}

\section{Problems}
\label{sec:problems}


\setcounter {probnum}{0}


\begin{problem}
In this and the next four problems, you will calculate
the one-loop counterterms for the $(\phi ^{3})_{6}$ theory, that is
$\phi ^{3}$ theory in space-time dimension 6.  When dimensional
regularization is used, so that the dimension of space
time is $n=6-\epsilon $, the Lagrangian is
\begin{eqnarray}
   {\cal L}&=& \BDpos \frac {1}{2}(\partial \phi )^{2} - \frac {1}{2}m^{2}\phi ^{2} -
\frac {\mu ^{\epsilon /2}g}{3!}\phi ^{3}
\nonumber\\
           & & \BDplus \frac {1}{2}\Delta Z(\partial \phi )^{2} - \frac {1}{2}\Delta
m^{2}\phi ^{2}
               - \frac { \mu ^{\epsilon /2}\Delta g}{3!}\phi ^{3} - \Delta
h\phi  .
\end{eqnarray}
Show that the counterterms indicated are all the ones
that are needed.  Note the presence of an additional
term $-\Delta h\phi $ linear in the field.
\end{problem}


\begin{problem}
Compute the values of all the one-loop counterterms. Assume that
the coupling renormalization $\Delta g$, the wave function
renormalization $\Delta Z$, and the mass renormalization $\Delta m^{2}$ are
defined by minimal subtraction. {\em But define the one-point
counterterm $\Delta h$ to cancel the tadpole graphs exactly, so that
$\langle 0|\phi |0\rangle = 0$.}
\end{problem}


\begin{problem}
What is the renormalization-group equation for this theory?  You
will need to show, by consideration of the vacuum expectation
value of the field, $\langle 0|\phi |0\rangle $, and its renormalization
condition,
that no term involving a coupling $h\phi $ is needed.
\end{problem}


\begin{problem}
Find the renormalization-group coefficients $\beta $, $\gamma $, and $ \gamma
_{m}$
for this theory.
\end{problem}


\begin{problem}
Still in $\phi ^{3}$ theory in six space-time dimensions, compute the
physical mass and the residue of the particle pole in the
propagator, both to order $g^{2}$.  Your results should be expressed
in terms of the renormalized mass and coupling in the MS scheme.
\end{problem}


\begin{problem}
The dependence of a Green function on the unit of mass $\mu $ in
each order of perturbation theory is a polynomial in $\ln\mu $.
Thus one can write the perturbation series as:
\begin{equation}
   \sum _{N} \lambda _{R}^{N}  \sum _{n=0}^{n_{\rm max}(N)} a_{Nn} (\ln\mu
)^{n} .
\end{equation}
(This polynomial dependence will follow from the methods you are
about to use.)  The degree of the polynomial is $n_{\rm max}(N)$,
and the subject of this and the next problem is to determine some
of its properties.  (Although the ideas are the same in any
theory, assume that we are working in $(\phi ^{4})_{4}$ theory.)

\noindent(a) Show from the renormalization group equation that
$n_{\rm max}(N+1) = 1+n_{\rm max}(N)$.

\noindent(b) Hence show that
$n_{\rm max}(N)=\mbox{number of loops}$, with the number of
loops being $N-1$ for the connected 4-point function.
\end{problem}


\begin{problem}
(a) With the same set up as in the previous problem, define the
leading coefficient in $N$th order to be $a_{Nn_{\rm max}}$.  Find a
recurrence relation between the leading coefficient at order
$N+1$ and at order $N$.  It will involve the one-loop coefficient
in the $\beta $ function.

\noindent(b) Hence (still in $(\phi ^{4})_{4}$ theory) find the value
of $a_{N,N-1}$ for the (connected) four-point Green function.
\end{problem}


\begin{problem}
Suppose that the bare coupling in $\phi ^{4}$ theory in two different
renormalization prescriptions is given by
\begin{equation}
   g_{0} = \mu ^{\epsilon } \left[
                g_{1}
                + g_{1}^{2} \left( \frac {A_{11}}{\epsilon } + A_{10}(\epsilon
) \right)
                + g_{1}^{3} \left( \frac {A_{22}}{\epsilon ^{2}}+\frac {
A_{21}}{\epsilon }+A_{20}(\epsilon ) \right)
                + \dots
            \right],
\end{equation}
\begin{equation}
   g_{0} = \mu '{}^{\epsilon } \left[
                    g_{2}
                    + g_{2}^{2} \left( \frac {B_{11}}{\epsilon } \right)
                    + g_{2}^{3} \left( \frac {B_{22}}{\epsilon ^{2}}+\frac
{B_{21}}{\epsilon } \right)
                    + \dots
                \right] .
\end{equation}
(Scheme 2 would be minimal subtraction.)  The coefficients $A_{10}$
and $A_{20}$ are functions of $\epsilon $, analytic at $\epsilon = 0$, and all
the
other coefficients are constants.

\noindent(a) Express the renormalized coupling, $g_{1}$, of scheme 1
as a power series in the renormalized coupling, $g_{2}$, of scheme
2.  You will be able to obtain the series up to the $g_{2}^{3}$ term
from the information given.

\noindent(b) What conditions must the coefficients satisfy in
order that the renormalized couplings both be finite.

\noindent(c) Show that in particular the 2-loop double
pole term ($A_{22}/\epsilon ^{2}$ or $B_{22}/\epsilon ^{2}$) is determined
completely
by the 1-loop single pole.
\end{problem}


\begin{problem}
Consider $\phi ^{4}$ theory (in 4 space-time dimensions).  Let $\lambda $
be the coupling when the minimal subtraction scheme
(MS) is used.  Let the coupling in another scheme be
related to it: $\lambda _{1}=\lambda _{1}(\lambda )$.

\noindent(a) What is the renormalization-group $\beta $ function
in this other scheme?  Call this function $\beta _{1}(\lambda _{1})$.  (The
answer should involve the function $\lambda _{1}(\lambda )$ and $\beta $ in the
MS
scheme.)

\noindent(b) Assume that $\lambda _{1}(\lambda )$ can be expanded in a
power series in $\lambda $, with first term $\lambda $.  Show that the
first two terms in $\beta _{1}(\lambda _{1})$ have the same coefficients as
in $\beta (\lambda )$.
\end{problem}


\begin{problem}
{\bf Hard problem! }

\noindent(a) Compute all the two-loop renormalization
coefficients in $\phi ^{4}$ theory.

\noindent(b) Hence show that
\begin{eqnarray*}
   \lambda _{0} &=& \mu ^{\epsilon } \left[
                \lambda  + \frac {3}{\epsilon } \frac { \lambda ^{2}}{16\pi
^{2}}
                + \frac {\lambda ^{3}}{(16\pi ^{2})^{2}} \left( \frac
{9}{\epsilon ^{ 2}} - \frac {17}{6\epsilon } \right)
                + O(\lambda ^{4})
             \right] ,
\\
   m_{0}^{2} &=& m^{2} \left[
                  1 + \frac {1}{\epsilon } \frac {\lambda }{16\pi ^{2}}
                  + \frac { \lambda ^{2}}{(16\pi ^{2})^{2}} \left( \frac
{2}{\epsilon ^{2}} - \frac {5}{12\epsilon } \right)
                  + O(\lambda ^{3})
              \right] ,
\\
   Z &=& 1 - \frac {1}{12\epsilon } \frac {\lambda ^{2}}{(16\pi ^{2})^{2}} +
O(\lambda ^{3}) .
\end{eqnarray*}
{\em Hints:}
\begin{itemize}

\item[i] If you do the integrals in the most obvious way they
    will be quite complicated.  But you can take advantage of the
    fact that, for example, the counterterms for the coupling are
    independent of the mass, and so set $m=0$ in the calculation
    of $\delta \lambda $.

\item[ii] Similarly, you may set $m=0$ in the self-energy when
    you are calculating $\delta Z$.

\item[iii] Also, when computing the divergent part of a graph
    to obtain $\delta \lambda $ take advantage of the fact that the overall
    divergence that you need is independent of the external
    momenta, and set some appropriate momenta to zero.

\end{itemize}
\end{problem}


\begin{problem}
The full Lagrangian with counterterms for QED with a
photon mass term is
\begin{eqnarray}
   {\cal L} &=& -\frac {1}{4}F_{\mu \nu }^{2} + \frac {1}{2}m^{2}A_{\mu }^{2} -
\frac {\lambda }{2} (\partial \cdot A)^{2}
      +i\bar{\psi } \st{\partial }\psi  + \mu ^{\epsilon /2}e\bar{\psi
}\st{A}\psi -M\bar{\psi }\psi
      \hspace*{0.6in}
\nonumber\\
   & & \hspace*{-0.35in}
       -\frac {Z_{3}-1}{4} F_{\mu \nu }^{2} + (Z_{2}-1)i\bar{\psi }\st{\partial
}\psi
       + (Z_{2}-1) \mu ^{\epsilon /2} e \bar{\psi }\st{A}\psi
       - (M_{0}Z_{2}-1)\bar{\psi }\psi  .
\label{eq:QED2}
\end{eqnarray}
Note that there are no counterterms for the photon
mass and gauge fixing terms.  This can be proved.

\noindent(a) What is the renormalization group equation for the
Green functions of the theory?

\noindent(b) From the renormalization counterterms (see
\cite{Sterman}),
derive the renormalization group coefficients ($\beta $, etc).

Note that the bare coupling and fields are defined by
\begin{eqnarray}
   e_{0} = \mu ^{\epsilon /2}eZ_{3}^{-1/2} ,
{}~~~
   A_{0} = AZ_{3}^{1/2} ,
{}~~~
   \psi _{0} = \psi Z_{2}^{1/2} ,
\end{eqnarray}
with corresponding definitions for the bare mass of the photon
and the bare gauge fixing parameter.  Note also that the total
variation with respect to $\mu $ is
\begin{equation}
   \mu \frac {d}{d\mu } = \mu \frac {\partial }{\partial \mu } +
          \mu \frac {d e}{d\mu } \frac {\partial }{\partial e} +
          \mu \frac {d M}{d\mu } \frac {\partial }{\partial M} +
          \mu \frac {d m^{2}}{d\mu } \frac {\partial }{\partial m^{2}} +
          \mu \frac {d\lambda }{d\mu } \frac {\partial }{\partial \lambda }.
\end{equation}
\end{problem}


\begin{problem}
In the theory of a combined $\phi ^{3}$ and $\phi ^{4}$ interaction
\begin{equation}
   {\cal L} = \BDpos \frac {1}{2}\partial \phi ^{2} - \frac {m^{2}}{2}\phi ^{2} -
\frac {g}{3!}\phi ^{3} - \frac {\lambda }{4!}\phi ^{4} ,
\label{phi34}
\end{equation}
find all the UV divergent one-loop graphs.  Which, if
any, counterterms are needed beyond those for terms in
Eq.~(\ref {phi34})?  Calculate the one-loop counterterm
for the three-point coupling $g$, using minimal subtraction
for the renormalization prescription.
\end{problem}


\begin{problem}
In this and the next problem, you will find the coordinate-space
propagator for a free scalar field of mass $m$ in an
$n$-dimensional space-time.  The propagator is
\begin{equation}
  i \Delta _{F}(x;n,m) = \int  \frac {d^{n}q}{(2\pi )^{n}} 
    \frac {i e^{\BDpos iq\cdot x}}{\BDpos q^{2}-m^{2}+i\epsilon } ,
\label{SFFT}
\end{equation}
and it satisfies the equation
\begin{equation}
   (\dalem +m^{2})\Delta _{F}=-\delta ^{(n)}(x) .
\label{defSF}
\end{equation}

\noindent(a) Now, $\Delta _{F}$ is a function of $x^{2}$ alone.  Write
\begin{equation}
   \Delta _{F} = (\BDneg x^{2})^{1/2-n/4} \, w \! \left(m\sqrt { \BDneg x^{2}}\right) ,
\end{equation}
and show that $w$ satisfies the modified Bessel equation of order
$\nu  = n/2-1$, when $x^{2} \not= 0$.  (You'll need to look up the
properties of such functions. Standard solutions are named $I_{\pm \nu }$
and $K_{\nu }$. The most general solution of the equation for $w$ is a
linear combination of two standard solutions of the modified
Bessel equation.  So some boundary conditions are needed.  One
boundary condition is that $\Delta _{F}\to 0$ as $\BDneg x^{2}\to \infty $, and the
other is
that the $\delta $-function in Eq.~(\ref{defSF}) be obtained.)

\noindent(b) To obtain the correct boundary condition at $x=0$,
Wick-rotate time ($x^{0}$) to be imaginary, integrate
Eq.~(\ref {defSF}) over the interior of a small hypersphere
centered at the origin, and use the divergence theorem.
Show that the leading singularity of $\Delta _{F}$ as $x\to 0$ is
\begin{equation}
   \Delta _{F} \sim \frac {-i \Gamma ( n/2 - 1 )}{4\pi ^{n/2} \, \left( \BDneg x^{2}
\right)^{n/2-1}} .
\end{equation}

\noindent(c) Now apply the two boundary conditions to
find a formula for $\Delta _{F}$ in terms of standard Bessel
functions, Eq.\ (\ref{eq:DeltaF}).
\end{problem}


\begin{problem}
In Minkowski space, the propagator $i\Delta _{F}(x^{2})$ is singular, not
only at $x=0$, but on the whole light-cone $x^{2}=0$. Given the $i\epsilon $
in the momentum-space propagator, as in Eq.~(\ref {SFFT}), what
is the correct $i\epsilon $ prescription in coordinate space?
You may find the book by Bogoliubov and Shirkov\cite{BS} helpful
for this and the previous problem.
\end{problem}





\end{document}